\documentclass[aps,preprint,epsf]{revtex4}
\usepackage{amsmath}
\usepackage{amsfonts}
\usepackage{graphicx,graphics}

\begin{document}

\title
{The (extended) dynamical mean field theory 
combined with the two-particle irreducible functional
renormalization-group\\
 approach as a tool to study strongly-correlated systems
}
\author{A. A. Katanin}
\address{Institute of Metal Physics of Ural Branch RAS, 620990, Kovalevskaya str. 18,
Ekaterinburg, Russia
}

\begin{abstract}
We propose new approach for treatment of local and non-local interactions
in correlated electronic systems, which 
uses self-energy and the two-particle irreducible
vertices, obtained from 
(extended) dynamical mean-field theory, 
as an input of two-particle
irreducible functional renormalization-group (2PI-fRG) approach. 
Using 2PI-fRG approach 
allows us to treat both, local and non-local interactions. 
In
case of purely local interaction the corresponding equations have similar (although not identical) structure to the earlier developed DMF$^2$RG approach.
In a
simplest truncation, 
neglecting scale-dependence of the two-particle irreducible vertices, we reproduce the results for the two-particle
vertices/susceptibilities in the ladder approximation of the dual boson or D$\Gamma$A approach; in more sophisticated truncations the method allows us to consider non-local corrections to the self-energy, as well as the interplay of charge- and spin correlations.
The proposed scheme is tested on the two-dimensional standard and extended $U$-$V$ half-filled Hubbard models.
For the standard Hubbard model we obtain non-local self-energy, which is in agreement with numerical studies; for the extended Hubbard model we obtain the boundary of charge instability, which agrees well with the results of the dual boson (DB) approach. We also find that the effect of spin correlations on electron interaction in the charge channel, not considered previously in the DB approach, 
only slightly reduces critical next-nearest-neighbor interaction of charge instability of the extended Hubbard model at the considered finite small temperature, yielding better agreement with dynamic cluster approximation. The considered method is rather general and can be applied to study
various phenomena in strongly-correlated electronic systems.
\end{abstract}

\maketitle


\section{Introduction}

Strongly-correlated electronic systems attract a lot of attention, since
they show a broad variety of interesting physical phenomena, such as spin-
and charge-density wave instabilities, as well as superconductivity,
originating from interelectron Coulomb interaction, see, e.g., Refs. \cite%
{Cr,Pnictide,CO1,High-Tc}. Screened part of this interaction can be
effectively described by the on-site or nearest-neighbor repulsion. 
Even local interactions, dressed by the particle-hole bubbles (e.g. in spin channel), yield attractive \textit{non-local }%
interaction in the superconducting channel \cite%
{KohnLuttinger,Scalapino}, as well as induce non-local interaction in the charge channel (see, e.g. Ref. \cite{Chubukov}). These effects can be further enhanced by non-local part of the interaction.

Developing suitable approximations for treatment of local and non-local
interactions in strongly-correlated electronic systems represents therefore an
important problem, since its solution allows to describe above mentioned physical phenomena, including also description of dynamic
screening in solids from first principles
(see, e.g., Refs.
\cite{GW,EDMFTGW,cRPA,AbInitioDGA}).
While local correlations, which appear due to the (non-)local interactions in strongly-correlated
systems are well described by the (extended) dynamical mean-field theory
((E)DMFT) \cite{DMFT,DMFT2,EDMFT,EDMFTGW,EDMFT_Si}, this theory is not
sufficient to describe the non-local correlations. The first step beyond
(E)DMFT was performed by (E)DMFT+GW approximation, proposed in Ref. \cite{EDMFTGW} to describe screening of Coulomb interaction in strongly correlated systems.
Recent progress in diagrammatic extensions of (E)DMFT \cite{Review}, namely dynamic vertex
approximation (D$\Gamma $A) \cite{DGA1a,DGA1b,DGA1c,DGA1d,DGA2,abinitioDGA},
dual fermion approach \cite{DF1,DF2,DF3,DF4,DF5}, and the dual boson (DB)
approach \cite{DB1,DB2,DB3,DB4} allowed to treat non-local correlations on a
non-perturbative basis. 
Yet, the conservation laws are fulfilled only in some special versions of these approaches (see, e.g., the discussion in Refs.  \cite{DB3,Review}).

The concept of $\Phi $-derivability, proposed long time ago \cite{BK},
allows to search for new approaches, which treat non-local correlations in
strongly-correlated systems. Although these approaches typically violate crossing symmetry, they may provide alternative view on correlated systems, which may fulfill better conservation laws. 
The fluctuation exchange approach (FLEX) \cite%
{Bickers,Bickers1} was proposed as a $\Phi $-derivable approximation, which can yield
self-energy and two-particle vertices, derived from the same functional, and
therefore fulfilling the conservation laws. However, this approach, being
perturbative, may not describe correctly properties of strongly-correlated
systems; the corresponding self-energy
contains only
the result of the summation of ladder diagrams with respect to the bare
interaction, and is not guaranteed to yield better results, than in the
other diagrammatic approaches, see, e.g., the discussion in Refs. \cite%
{Tremblay,VariationalQMC}. The recently proposed TRILEX approach \cite%
{TRILEX} extends the concept of $\Phi $-derivability to merge dynamical
mean-field theory with perturbation techniques; this approach is however restricted to the approximation of locality of the three-point (fermion-boson) vertices.

Recently, the two-particle irreducible functional renormalization-group
(2PI-fRG) approach based on considering the evolution of the Luttinger-Ward functional $\Phi$ with some parameter (e.g. switching on the interaction) was proposed \cite{Dupuis,Dupuis1} and its application to
quantum anharmonic oscillator \cite{Meden} and single-impurity Anderson
model \cite{Meden1} was discussed and suitable truncation schemes were
developed. For strongly-correlated systems, however, the standard
truncations, applied within this approach, may not be sufficient, which
makes important search for non-perturbative starting points of $\Phi $. In
this respect, the (E)DMFT provides a natural starting point for the search
of new functionals for strongly-correlated systems. In the present paper we
propose the scheme to merge of (E)DMFT and the 2PI-fRG approach. The
suggested scheme follows earlier considered DMF$^{2}$RG approach\cite%
{DMF2RG,DMF2RG2,DMF2RG3}, which merges DMFT with one-particle irreducible (1PI)
functional renormalization group \cite{fRGReview} by using information from
the DMFT (the self-energy and one-particle irreducible vertices) as a
starting point for the fRG flow. The treatment of the non-local interactions require, however, considering two-particle irreducible vertices, since the interaction can not contain scale dependence in the standard applications of the 1PI fRG method. 

The approach, considered in the present paper,
uses two-particle irreducible vertices, which 
allows us to treat the non-local interactions beyond (E)DMFT.
Although it was shown recently that in
the strong-coupling regime the charge and superconducting 2PI vertices may
be singular \cite{Toschi,div1,div2} (which is related to the problem
of $\Phi \lbrack G]$ being not uniquely defined, cf. Ref. \cite{BKFail,BKFail1}),
this problem may be relevant for considering flow of the two-particle vertices only at sufficiently strong coupling in the local
moment regime \cite{Toschi,div2}; 
in some cases this problem
can be circumvented by an appropriate treatment of the corresponding
channels. 
We formulate
the (E)DMFT+2PI-fRG method, relate it to known approaches to
strongly-correlated systems and test its applicability to study charge instability in the half filled two dimensional $U$-$V$ model. The plan of the paper is the following. In
Sect. II we introduce the model and formulate the (E)DMFT approach in the
notations, suitable for the following discussion. In Sect. III we describe
the (E)DMFT+2PI-fRG approach and derive the respective equations. In Sect. IV we apply the developed approach to 
the half filled two dimensional standard and extended $U$-$V$ Hubbard model. In Sect. V we present conclusions
and discuss perspectives of the presented approach.

\section{The model and extended dynamical mean-field theory}

We consider a general one-band model of interacting fermions 
\begin{equation}
H=\sum\limits_{\mathbf{k},\sigma }\varepsilon _{\mathbf{k} }\widehat{c%
}_{\mathbf{k},\sigma }^{+}\widehat{c}_{\mathbf{k},\sigma }+H_{\mathrm{int}}[%
\widehat{c},\widehat{c}^{+}],  \label{H}
\end{equation}%
where $\widehat{c}_{i\sigma },\widehat{c}_{i\sigma }^{+}$ are the fermionic
operators, and $\widehat{c}_{\mathbf{k},\sigma },\widehat{c}_{\mathbf{k}%
,\sigma }^{+}$ are their Fourier transforms, $\sigma =\uparrow ,\downarrow $
corresponds to a spin index. The interaction $H_{\mathrm{int}}$ contains in
general both, local $U$ and non-local $V_{ij}^{c(s)}$ contributions, the
latter act on charge and spin degrees of freedom, 
\begin{equation}
H_{\mathrm{int}}[\widehat{c},\widehat{c}^{+}]=\sum\limits_{i}Un_{i\uparrow
}n_{i\downarrow }+\frac{1}{2}\sum\limits_{ij}\left(
V_{ij}^{c}n_{i}n_{j}+V_{ij}^{s}\mathbf{S}_{i}\mathbf{S}_{j}\right) ,
\label{Hint}
\end{equation}%
where $n_{i\sigma }=\widehat{c}_{i\sigma }^{+}\widehat{c}_{i\sigma }$, $n_i=\sum_\sigma n_{i\sigma }$,  and %
$\mathbf{S}_{i}=\sum_{\sigma,\sigma'}\widehat{c}_{i\sigma }^{+}\mbox {\boldmath $\sigma $}_{\sigma \sigma
^{\prime }}\widehat{c}_{i\sigma ^{\prime }},$ $\mbox {\boldmath $\sigma $}_{\sigma
\sigma ^{\prime }}$ are the Pauli matrices, $V^{c(s)}_{ij}$ depends on the distance between sites $i$ and $j$ only.

The model is characterized by the generating functional%
\begin{eqnarray}
Z[\eta ,\eta ^{+}] &=&\int d[c,c^{+}]\exp \left\{ -\mathcal{S}[c,c^{+}]+\eta
^{+}c+c^{+}\eta \right\},   \label{gen} \\
\mathcal{S}[c,c^{+}] &=&\int d\tau \left\{ \sum\limits_{i,\sigma }c_{i\sigma
}^{\dagger }(\tau )\frac{\partial }{\partial \tau }c_{i\sigma }(\tau
)+H[c,c^{+}]\right\} ,
\end{eqnarray}%
where $c_{i\sigma },c_{i\sigma }^{+},\eta _{i\sigma },\eta _{i\sigma }^{+}$
are the Grassmann fields, the fields $\eta _{i\sigma },\eta _{i\sigma }^{+}$
correspond to source terms, $\tau \in \lbrack 0,\beta =1/T]$ is the
imaginary time, $T$ is the temperature. The (extended) dynamical mean-field theory \cite%
{EDMFTGW,EDMFT_Si,EDMFT} for the model (\ref{H}) can be introduced via the
corresponding local action 
\begin{equation}
\mathcal{S}_{\mathrm{(E)DMFT}}[c,c^{+}]=
-\sum\limits_{k,\sigma }c_{k\sigma }^{+ }\zeta ^{-1}(i \nu_n
)c_{k\sigma }+\mathcal{S}_{\mathrm{int}}^{\mathrm{loc}}[c_{i\sigma
},c_{i\sigma }^{+}],  \label{VDMFT}
\end{equation}%
where 
\begin{equation}
\mathcal{S}_{\mathrm{int}}^{\mathrm{loc}}[c_{i\sigma },c_{i\sigma
}^{+}]=U\sum\limits_{q} n_{q\uparrow }n_{-q,\downarrow }+\frac{1}{2}%
\sum\limits_{q}\left[ v^{c}(i\omega _{n})n_{q}n_{-q}+v^{s}(i\omega _{n})%
\mathbf{S}_{q}\mathbf{S}_{-q}\right] ,
\label{HintEDMFT}
\end{equation}%
$c_{k\sigma}=
\int d\tau \sum_i c_{i\sigma}\exp(i\omega_n\tau-i{\bf kR}_i)$, 
$n_q=\sum_\sigma {n_{q\sigma}}=\sum_{k\sigma} c^{+}_{k\sigma} c_{k+q,\sigma}$, and ${\bf S}_q=\sum_{k\sigma} c^{+}_{k\sigma} \mbox {\boldmath $\sigma $}_{\sigma\sigma'}c_{k+q,\sigma'}$ are the Fourier transforms of the respective quantities
(we use the $4$-vector notation $k=(\mathbf{k}%
,i\nu _{n})$, $q=(\mathbf{q}%
,i\omega _{n})$ and assume factor of $T$ for every frequency summation);
the ``Weiss field" functions $\zeta (i\nu _{n})$ and $v(i\omega _{n})$ have
to be determined self-consistently from the conditions 
\begin{subequations}
\label{SC} 
\begin{eqnarray}
G_{\mathrm{loc}}(i\nu _{n}) &\equiv &\frac{1}{\zeta ^{-1}(i\nu _{n})-\Sigma
_{\mathrm{loc}}(i\nu _{n})}=\sum\limits_{\mathbf{k}}
\frac{1}{
G_{0,k}^{-1}-\Sigma _{\mathrm{loc}}(i\nu _{n})},
\label{sc} \\
\chi _{\mathrm{loc}}^{\mathrm{c(s)}}(i\omega _{n}) &\equiv &\frac{1}{%
v^{c(s)}(i\omega _{n})+\Pi _{\mathrm{loc}}^{\mathrm{c(s)}}(i\omega _{n})}%
=\sum\limits_{\mathbf{q}}\chi _{\rm EDMFT}^{\mathrm{c(s)}}(\mathbf{q},i\omega _{n}),
\label{sc1}
\end{eqnarray}%
\end{subequations}
where 
\begin{eqnarray}
\chi_{\rm EDMFT} ^{\mathrm{c(s)}}(\mathbf{q},i\omega _{n}) &\equiv &\chi _{{\rm EDMFT}, q}^{\mathrm{c(s)}}=%
\left[ V_{\mathbf{q}}^{c(s)}+\Pi _{\mathrm{loc}}^{c(s)}(i\omega _{n})\right]
^{-1}  \nonumber \\
&=&\left[ (\chi _{\mathrm{loc}}^{\mathrm{c(s)}}(i\omega
_{n}))^{-1}-v^{c(s)}(i\omega _{n})+V_{\mathbf{q}}^{c(s)}\right] ^{-1},
\label{chiq_EDMFT}
\end{eqnarray}%
$G_{0,k}^{-1}=i\nu _{n}-\varepsilon _{\mathbf{k}}+\mu_{\rm loc}$ is the lattice
noninteracting Green function, $\mu_{\rm loc}$ is the (E)DMFT chemical potential, $V_{\mathbf{q}}^{c(s)}$ are
the Fourier transformed\ interactions $V_{ij}^{c(s)},$ and $\Sigma _{\mathrm{%
loc}}(i\nu _{n})$ and $\Pi _{\mathrm{loc}}^{c(s)}(i\omega _{n})$ are the
fermionic and bosonic self-energy of the impurity problem (\ref{VDMFT}),
which is in practice obtained within one of the impurity solvers: exact
diagonalization, quantum Monte-Carlo (QMC), etc. These solvers provide
information not only on the electronic self-energy, but also the
corresponding vertex functions\cite{DGA1a,Toschi1} 
\begin{eqnarray}
{\mathcal F}_{\mathrm{loc}}^{\sigma \sigma ^{\prime }}(i\nu _{1}...i\nu _{3})
&=&(1+\delta _{\sigma \sigma ^{\prime }})^{-1}G_{%
\mathrm{loc}}^{-1}(i\nu _1+i\nu_2-i\nu_3)\prod\nolimits_{i=1}^{3}G_{%
\mathrm{loc}}^{-1}(i\nu _{i})  \label{Gamma_loc} \\
&&\times \left[ G_{\mathrm{loc,}\sigma \sigma ^{\prime }}^{(4)}(i\nu
_{1}...i\nu _{3})-G_{\mathrm{loc}}(i\nu _{1})G_{\mathrm{loc}}(i\nu
_{2})(\delta _{\nu _{1}\nu _{3}}-\delta _{\sigma \sigma ^{\prime }}\delta
_{\nu _{2}\nu _{3}})\right] ,  \nonumber
\end{eqnarray}%
$G_{\mathrm{loc}}^{(4)}$ is the two-particle local Green function, which can
be obtained via the solution of the impurity problem. Solving Bethe-Salpeter
equations in the spin- and charge channel then provides an information about
the respective two-particle irreducible vertices in spin and charge
channels, $\Phi _{\mathrm{loc}}^{(2),c(s)}(i\nu _{1}...i\nu _{3}),$ see Refs. 
\cite{DGA1a,Toschi1,Review}.

\section{The two-particle irreducible functional renormalization-group
approach}

\subsection{General formalism}

\begin{figure}[tbp]
\center \includegraphics[width=0.7 \linewidth]{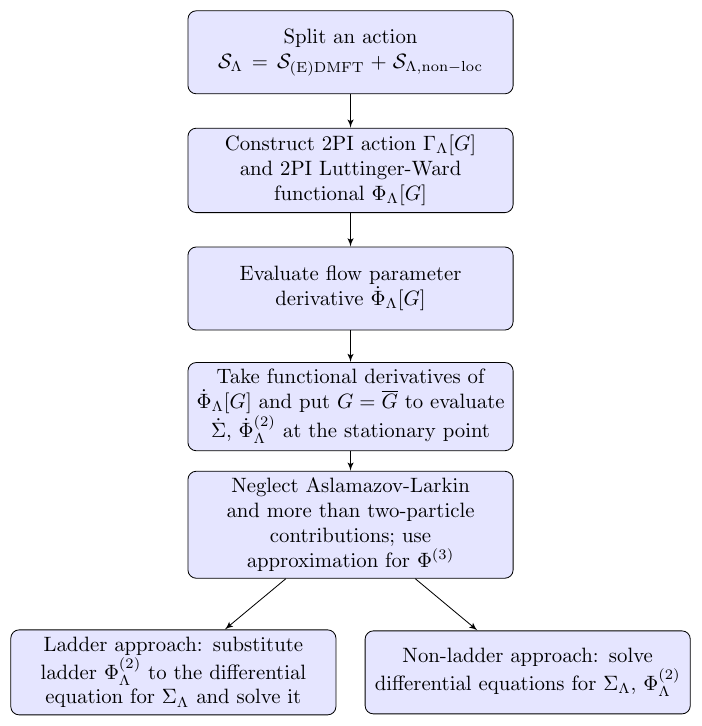} 
\endcenter
\caption{(Color online) The flow diagram of derivation and solution of 2PI equations}
\label{FigFlow}
\end{figure}

The considering approach is similar to the DMF$^{2}$RG approach for the flow
from infinite to finite number of dimensions for the standard Hubbard model \cite{DMF2RG}, but follows the 2PI fRG approach (see, e.g., Ref. \cite{Dupuis} and the schematic flowchart in Fig. \ref{FigFlow}). In particular, we consider the evolution of generating
functional with the action 
\begin{equation}
\mathcal{S}_{\Lambda }=\mathcal{S}_{\mathrm{(E)DMFT}}[c,c^{+}]+\mathcal{S}_{\Lambda,\text{%
non-loc}}[c,c^{+}],  \label{S_L}
\end{equation}%
where%
\begin{eqnarray}
\mathcal{S}_{\Lambda,\text{non-loc}}[c,c^{+}]&=&\Lambda
\sum_{k\sigma}c_{k\sigma }^{+}\left[\zeta ^{-1}(i\nu _{n})-G_{0,k}^{-1}\right]c_{k\sigma }-
\left(\mu_\Lambda-\mu_{\rm loc}\right)\sum_{k\sigma}c_{k\sigma }^{+}
c_{k\sigma }  \nonumber \\
&+&\frac{\Lambda 
}{2}\sum_{q}\left( \widetilde{V}_{q}^{c}n_{q}n_{-q}+%
\widetilde{V}_{q}^{s}\mathbf{S}_{q}\mathbf{S}_{-q}\right) ,
\label{Snonloc}
\end{eqnarray}%
and $\widetilde{V}_{q}^{c(s)}=V_{\mathbf{q}}^{c(s)}-v^{c(s)}(i\omega
_{n})$, $\mu_\Lambda$ is the scale-dependent chemical potential, which can be determined, e.g., from the condition of constant number of particles during the flow. For $\Lambda =0$ the (E)DMFT theory is reproduced,
while for $%
\Lambda =1$ we obtain the lattice problem (\ref{gen}). Note that the $\Lambda$-dependence of the interaction part in Eq. (\ref{S_L}), which is non-multiplicative because of the part of the interaction, contained in $\mathcal{S}_{\mathrm{(E)DMFT}}$, prevents reducing the action $\mathcal{S}_\Lambda$ to that containing $\Lambda$ dependence of the quadratic part only (cf. Ref. \cite{HonerkampU}), which makes it difficult using the 1PI fRG, e.g.  DMF$^2$RG \cite{DMF2RG,DMF2RG3} approach at finite $\widetilde V^{c(s)}$. Although 1PI approach can be in principle generalized to include arbitrary $\Lambda$-dependence of the interaction, the structure of the resulting hierarchy of the equations is expected to be rather complicated, and not considered here.

As discussed above, we follow instead the 2PI approach \cite{Dupuis,Meden}, which 
considers the evolution of the partition function%
\begin{equation}
Z_{\Lambda }[J]=\int D[c,c^{+}]
\exp \left[
-\mathcal{S}_{\Lambda
}+2\sum\limits_{k,q;m=c,s_a}J_{k,q}^{m} M_{k,k+q}^{m_+}+\sum\limits_{k,q;m={\rm ssc}_{\pm},{\rm tsc}_{a,\pm}}J_{k,q}^{m} M_{k,-k+q}^{m}\right],  \label{ZJ}
\end{equation}%
where $J^{m}$ are the source fields, $M_{k,k'}^{m_r}=\sum\nolimits_{r',\sigma\sigma'
}C_{rk\sigma }^{+}{\mathcal M}^{m}_{r\sigma,r'\sigma'} C_{r'k'\sigma' }$ ($r,r'=+,- $)
are
the respective combinations of Grassmann variables $C_{+,k\sigma}=c_{k\sigma}$, $C_{-,k\sigma}=c^+_{k\sigma}$, corresponding to charge $\mathcal{M}_{r\sigma, r'\sigma'}^{c}=r\delta_{\sigma\sigma'}\delta_{rr'}/2$, spin
$\mathcal{M}_{r\sigma, r'\sigma'}^{s_a}=({\sigma }^a_{\sigma \sigma ^{\prime
}}\delta_{r,+}-{\sigma }^a_{\sigma' \sigma }\delta_{r,-})\delta_{rr'}/2$, singlet pairing $\mathcal{M}_{r\sigma,r'\sigma'}^{\mathrm{ssc}_\pm}=i r\sigma _{\sigma \sigma ^{\prime
}}^{y}\delta_{r,\pm}\delta_{-r,r'}$, and triplet pairing $\mathcal{M}_{r\sigma,r'\sigma'}^{\mathrm{tsc}_{a,\pm}}=i({\sigma }^a%
\sigma ^{y})_{\sigma \sigma ^{\prime }}\delta_{r,\pm}\delta_{-r,r'}$, $a=x,y,z$.
Performing Legendre
transform, we obtain the 2PI effective action 
\begin{equation}
\Gamma _{\Lambda }[G]=-\ln Z_{\Lambda }[J]+\sum\limits_{k,q,m}
J^{m}_{k,q} G^{m}_{k,\pm k+q}
,  \label{GammaG}
\end{equation}%
here and below the sums over $m$ are taken over $c,s_a,{\rm ssc}_\pm,{\rm tsc}_{a,\pm}$, if not specified differently, $+$ sign corresponds to the particle-hole ($m=c,s_a$), while $-$ to the particle-particle ($m={\rm ssc}_\pm,{\rm tsc}_{a,\pm}$) order parameters,
we also assume $\widetilde{V}^{s_a}_q=\widetilde{V}^{s}_q$, $\widetilde{V}^{{\rm ssc}_\pm,{\rm tsc}_{a,\pm}}_q=0$ for uniformity of notations. The requirement that $\Gamma_\Lambda[G]$ does not depend explicitly on $J$ yields $G^m_{k,\pm k+q}=\delta \ln Z_\Lambda[J]/\delta J_{k,q}^m$.

Following the
standard strategy, we introduce the Luttinger-Ward functional $\Phi_\Lambda[G]$ by \cite{BK} 
\begin{equation}
\Gamma _{\Lambda }[G]=\frac{1}{2}\mathrm{Tr}\ln \left( -\widehat{G}\right) +%
\frac{1}{2}\mathrm{Tr}\left[ I-\widehat{G}_{0\Lambda }^{-1} \widehat{G}\right]
-\Phi _{\Lambda }[G],
\end{equation}%
where 
\begin{eqnarray}
&&\widehat{G}_{0\Lambda ,kr\sigma,k^{\prime
}r'\sigma'}^{-1}=G_{0\Lambda,k}^{-1}\delta _{kk^{\prime }}\mathcal{M}^c_{r\sigma,r'\sigma'},\notag\\
&&G_{0\Lambda ,k}^{-1}=(1-\Lambda )\zeta ^{-1}(i\nu _{n})+\Lambda G_{0,k}^{-1}+\mu_\Lambda-\mu_{\rm loc},
\label{G0Lk}\\
%
&&\widehat{G}_{kr\sigma,k^{\prime },r'\sigma'}=\sum\limits_{m=c,s_a}\left(G_{k,k^{\prime }}^m \delta_{r,+}+G_{k',k}^m \delta_{r,-}\right)%
\mathcal{M}_{r\sigma,r^{\prime }\sigma'}^m +\sum\limits_{m={\rm ssc}_\pm,{\rm tsc}_{a,\pm}}G_{k,k^{\prime }}^m%
\mathcal{M}_{r\sigma,r^{\prime }\sigma'}^m,
\end{eqnarray}
$\mathrm{Tr}$ is taken with respect to fermionic momenta,
frequency, spin, and $r$ indices, and matrix multiplication of quantities with "hat" with respect to $k,r,\sigma$ indexes is assumed. 

Taking the derivative with respect to $G$, we obtain
\begin{eqnarray}
\Gamma _{\Lambda ,k,\pm k+q}^{(1),m} &\equiv &\frac{\delta \Gamma_\Lambda }{\delta
G_{k,\pm k+q}^m}=\frac{1}{2}\mathrm{Tr}\left[ \widehat{G}^{-1}\widehat{M}^{mkq}
\right] -G_{0\Lambda ,k}^{-1}\delta _{m,c}\delta_{q0}-\frac{\delta \Phi_\Lambda }{%
\delta G_{k,\pm k+q}^m}  
=J_{k,q}^m,
\label{Gamma_first}
\end{eqnarray}%
where $\widehat{M}^{c(s_a),KQ}_{kr\sigma,k'r'\sigma'}=(\delta_{kK}\delta_{k',K+Q}\delta_{r,+}+\delta_{k,K+Q}\delta_{k',K}\delta_{r,-})\mathcal{M}^{c(s_a)}_{r\sigma,r'\sigma'}$ and $\widehat{M}^{{\rm ssc}_\pm ({\rm tsc}_{a,\pm}),KQ}_{kr\sigma,k'r'\sigma'}=\delta_{kK}\delta_{k',-K+Q}\mathcal{M}^{{\rm ssc}_\pm ({\rm tsc}_{a,\pm})}_{r\sigma,r'\sigma'}$.
The stationary one- and two-particle quantities
(which we denote by bar) are determined by the condition $\Gamma ^{(1)}=0$,
such that 
\begin{eqnarray}
\overline{G}_{\Lambda ,k}^{-1} &\equiv &2(\overline{G}_{\Lambda
,k}^c)^{-1}=G_{0\Lambda ,k}^{-1}-\Sigma _{\Lambda ,k},  \nonumber \\
\Sigma _{\Lambda ,k} &=&-\overline{\Phi }_{\Lambda ,k}^{(1),c}  \label{GG}
\end{eqnarray}%
(we denote by $\Phi^{(n),m}_\Lambda$ the derivatives $\delta^n \Phi_\Lambda/\delta (G^m)^n$, e.g. $\Phi^{(1),c}_{\Lambda,k}=\delta \Phi_\Lambda/\delta G_{k}^c$). At $\Lambda =0$ we have $\Phi _{\Lambda =0}[G] =\Phi _{\mathrm{loc}}[G]$, such that
\begin{subequations}
\label{InitCond}
\begin{eqnarray}
\overline{\Phi }_{\Lambda =0,k}^{(1),c} &=&-\Sigma _{\Lambda =0,k}=-\Sigma _{%
\mathrm{loc}}(i\nu _{n}),
\label{Initb}\\
\overline{\Phi }_{\Lambda =0,kk'q}^{(2),mm^{\prime }} &=&\Phi _{\mathrm{loc},\nu\nu'\omega%
}^{(2),m}\delta _{mm^{\prime }},
\label{Initc}
\end{eqnarray}%
\end{subequations}
in the equation (\ref{Initc}) $k,k'$, and $q$ correspond to the initial, final momentum and momentum transfer. To keep the initial ($\Lambda=0$) Green function equal to its (E)DMFT value, we also choose the initial chemical potential $\mu_{\Lambda=0}=\mu_{\rm loc}$.
From this setup
we obtain the following 2PI-fRG equation (see Appendix A, cf. Refs. \cite%
{Dupuis,Meden}):%
\begin{eqnarray}
\dot{\Phi}_{\Lambda }[G] &=&\frac{1}{2}\sum_{kk^{\prime }q,m}\widetilde{V}%
_{q}^{m}\left\{ \left[ \Pi ^{-1}+\Phi _{\Lambda }^{(2)}
\right]
_{kk^{\prime }q,mm}^{-1}-G_{k,k+q}^m G_{k^{\prime },k'-q}^m \right\},  
\label{PhiRG}
\end{eqnarray}%
where $[\Pi _{kk^{\prime }q}^{mm^{\prime }}]^{-1}=(1/2)\mathrm{Tr}%
\left[ \widehat{G}
^{-1}\widehat{M}^{m',\pm k'+q,-q} 
\widehat{G}
^{-1}
\widehat{M}^{m,\pm k,q} 
\right]
$ is the inverse polarization bubble (cf. Refs. \cite{Dupuis,Meden}), the
inversion is performed with respect to momentum $k,k^{\prime }$ and channel $%
m,m^{\prime }$ indices. 
Taking functional derivatives and considering stationary
quantities, allowing to determine the self-energy and the two-particle vertex, which are represented as
\begin{eqnarray}
\Sigma_\Lambda&=&\widetilde \Sigma _{\Lambda}+2\Lambda ({V}_{{\bf q}=0}^{c}-v^c (0))\sum_{k^{\prime }} {\overline G}_{\Lambda,k^{\prime
}},\notag \\
\overline\Phi _{\Lambda,kk^{\prime }q}^{(2),c(s)} &=&\widetilde\Phi _{\Lambda,kk^{\prime }q}^{(2),c(s)}-\Lambda ({V}_{\bf q}^{c(s)}-v^{c(s)}(i\omega_n)),
\end{eqnarray}
neglecting the contribution of more than two-particle processes, as well as Aslamazov-Larkin contributions,
we find the equations for the self-energy and the 2PI vertices 
(see Appendix~A) 
\begin{subequations}
\begin{eqnarray}
\frac{d \widetilde \Sigma _{\Lambda ,k}}{d\Lambda } &=&-2\sum\limits_{q,m=c,s}a_{m}^{(c)}%
\left(\widetilde{\partial }_{\Lambda }{\mathcal F}_{\Lambda ,kkq}^{m}\right)\overline{G}_{\Lambda
,k+q}-2\sum\limits_{k^{\prime }}\widetilde{\Phi }_{\Lambda,kk^{\prime }0}^{(2),c}%
\frac{{d}\overline{G}_{\Lambda ,k^{\prime }}}{{d}\Lambda },  \label{Sflow_eq}
\\
\frac{d\widetilde{\Phi }_{\Lambda ,kk^{\prime }q}^{(2),c(s)}}{d\Lambda }
&=&\sum\limits_{m=c,s}a_{m}^{(c(s))}\widetilde{\partial }_{\Lambda
}{\mathcal F}_{\Lambda ,k,k+q,k^{\prime }-k}^{m}  \nonumber \\
&&+\sum\limits_{p,m=c,s}a_{m}^{(c(s))}{\mathcal F}_{\Lambda,k,p,k'-k }^{m} \frac{d \Pi^{m}_{\Lambda,p,k'-k} }%
{{d}\Lambda }
{\mathcal F}_{\Lambda,p,k+q,k'-k
}^{m},  \nonumber \\
&&+\frac{1}{2}\sum\limits_{p,m=\mathrm{ssc},\mathrm{tsc}}b_{m}^{(c(s))}{\mathcal F}_{\Lambda,k,p,k'+k+q }^{m} \frac{d \Pi^m_{p,k'+k+q}}%
{{d}\Lambda }
{\mathcal F}_{\Lambda,p,k+q,k'+k+q
}^{m},
\label{ifflow_eq}
\end{eqnarray}%
\label{flow_eq} 
\end{subequations}
where the coefficients in the charge- and spin
channels are given by $%
a_{c}^{(c)}=a_{c}^{(s)}=-a_{s}^{(s)}
=1/2, 
$ $a_{s}^{(c)}
=3/2,$ $b_{\rm tsc}^{(c)}=3$, $b^{(c)}_{\rm ssc}=b^{(s)}_{\rm tsc}=-b^{(s)}_{\rm ssc}=1$,
\begin{subequations}
\begin{eqnarray}
\widetilde{\partial }_{\Lambda }{\mathcal F}_{\Lambda ,kk^{\prime }q}^{m} &=&\left[ 
\overline{\Pi }_{\Lambda }^{-1}\overline{W}_{\Lambda }^{(2)}\widetilde{V}%
_{q}^{m}\overline{W}_{\Lambda }^{(2)}\overline{\Pi }_{\Lambda }^{-1}\right]
_{kk^{\prime }q,mm}, \\
\overline{W}_{\Lambda ,kk^{\prime }q}^{(2),mm^{\prime }} &=&\left[ \overline{%
\Pi }_{\Lambda }^{-1}+\overline{\Phi }_{\Lambda }^{(2)}
\right]
_{kk^{\prime }q,mm^{\prime }}^{-1}, \label{WVert}\\
{\mathcal F}_{\Lambda ,kk^{\prime }q}^{m
} &=&
\left[ \overline{%
\Pi }_{\Lambda }+\left(\overline{\Phi }_{\Lambda }^{(2)}
\right)^{-1}
\right]
_{kk^{\prime }q,mm
}
^{-1}
, \label{GammaVert1}\\
\overline{\Pi }_{\Lambda ,kk^{\prime }q}^{mm^{\prime }} &=&\Pi ^m_{\Lambda,k,q}\delta _{kk^{\prime }}\delta _{mm^{\prime }}\ 
\end{eqnarray}%
\end{subequations}
(we omit momenta/frequency indices in the matrix products with respect to $%
k,k^{\prime }$); $\Pi^{c(s)} _{\Lambda,k,q}=2%
\overline{G}_{\Lambda ,k}\overline{G}_{\Lambda ,k+q}$ and $\Pi^{\rm ssc(tsc)} _{\Lambda,k,q}=%
\overline{G}_{\Lambda ,k}\overline{G}_{\Lambda ,-k+q}$, here and in the following we denote $s:=s_z$, ${\rm ssc}:={\rm ssc}_+$, ${\rm tsc}:={\rm tsc}_{z,+}$ (the other components are accounted via the coefficients $a^{(c(s))}_m$ and $b^{(c(s))}_m$, reflecting the SU(2) invariance of the model).
Eq. (\ref{GammaVert1}) represents the Bethe-Salpeter equation for the {\it one-particle irreducible} vertices ${\mathcal F}_{\Lambda ,kk^{\prime }q}^{c(s)}$; the quantities $\widetilde{\partial }_{\Lambda
}{\mathcal F}_{\Lambda }^{(2),c(s)}$ can be represented as a certain scale derivative, $%
\widetilde{\partial }_{\Lambda }=\widetilde V^{c(s)}
\partial _{\overline \Phi_{\Lambda }^{c(s)}}$ of these vertices. %

\begin{figure}[tbp]
\center \includegraphics{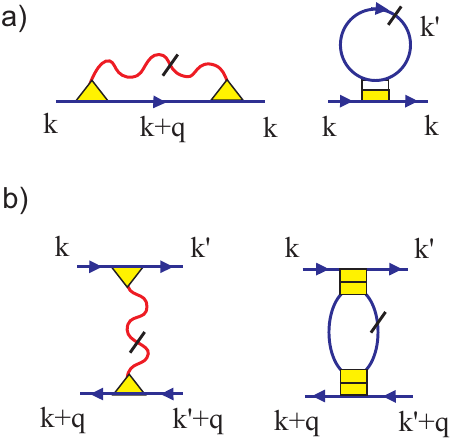} \endcenter
\caption{(Color online) The diagrammatic form of the right hand sides of Eqs. (\ref{flow_eq}) for the non-local
self-energy $\dot{\widetilde \Sigma}_{\Lambda,k}$ (a) and the two-particle irreducible vertex $%
\dot{\widetilde \Phi}^{(2),c(s)}_{\Lambda,kk'q}$ (b). Wavy lines with dash and triangular vertices
correspond to the scale derivative of the two-particle interaction $\widetilde{\partial }_{\Lambda }{\mathcal F}^{m}_{\Lambda}$, or,
equivalentely, $\protect\gamma^m {\widetilde{\partial }_{\Lambda }W}^m_{\Lambda} \protect\gamma^m$
(see text), solid lines correspond to the stationary Green functions $%
\overline{G}_{\Lambda }$, the half filled square box - to the two-particle
irreducible vertices $\widetilde \Phi^{(2),m}_{\Lambda}$, the filled square box - to the 1PI vertex $\mathcal{F}^{m}_{\Lambda}$. The line inside the boxes for $\mathcal{F}$ and $\widetilde \Phi$ is perpendicular to their direction of their two-particle reducibility and irreducibility, respectively. The dash on the solid lines corresponds to the derivative $d/d{\Lambda }$.}
\label{FigDiagr}
\end{figure}

The diagrammatic form of Eqs. (\ref{flow_eq}) is shown in Fig. \ref{FigDiagr}. The terms in the
right-hand side describe contributions from the non-local interaction (first
term) and local interaction (other terms), which are mixed during fRG flow.
Second and third terms in the right-hand side of equation
(\ref{ifflow_eq}) 
generate fRG analogue of the parquet diagrams with the
bare 2PI interactions. The solution of Eq. (\ref{ifflow_eq})
corresponds to renormalization of the bare vertices $\Phi _{\Lambda=0}^{(2),c(s)}$
by non-local interaction, as well by particle-hole and particle-particle bubbles, in the ``transverse" direction
with respect to that, in which the vector of the transfer momentum and frequency $q$ of $\overline{\Phi }^{(2),c(s)}_{\Lambda,kk'q}$ is
defined (see Fig. \ref{FigDiagr}b). 



Substituting the bare local interaction  $\widetilde \Phi _{\Lambda=0}^{(2),c(s)}=\Phi _{\rm loc}^{(2),c(s)}$ to the right-hand side of the equation (\ref{ifflow_eq}), the result of the integration over $\Lambda$, which corresponds to first iteration towards the solution of the differential equation, can be obtained analytically and represents all possible ladder diagrams in the ``transverse" channel:
\begin{eqnarray}
{\widetilde{\Phi }_{\Lambda ,kk^{\prime }q}^{(2),c(s)}}
=\Phi _{\mathrm{loc},\nu \nu'\omega}^{(2),c(s)}&
-&\sum\limits_{m=c,s}a_{m}^{(c(s))}\left[{\mathcal F}_{\Lambda,\nu,\nu+\omega,k'-k }^{ m}-{\mathcal F}_{{\Lambda=0},\nu,\nu+\omega,\nu'-\nu }^{m}\right]_{\widetilde \Phi_\Lambda=\Phi_{\rm loc
}}\notag \\
&-&\frac{1}{2}\sum\limits_{m={\rm ssc,tsc}}b_{m}^{(c(s))}\left[{\mathcal F}_{\Lambda,\nu,\nu+\omega,k'+k+q }^{m}-{\mathcal F}_{{\Lambda=0},\nu,\nu+\omega,\nu'+\nu+\omega }^{m} \right]_{\widetilde \Phi_\Lambda=\Phi_{\rm loc
}}. \label{Phi21st}
\end{eqnarray}
Note that due to using the bare vertices $\Phi_{\rm loc}$, the vertices $\mathcal F^m_\Lambda$ in the square brackets depend on momenta via the respective momentum transfers ($\bf k'-k$ for $m=c(s)$ and $\bf k'+k+q$ for $m={\rm ssc,tsc}$) only.  Subtraction of the $\Lambda=0$ vertices (i.e. their local part) in Eq. (\ref{Phi21st}) makes these vertices particle-hole irreducible in the ``longitudinal" direction. The approximation (\ref{Phi21st}) is referred below as the ``transverse channel ladder approximation".

In general, the equations (\ref{flow_eq}) should be solved numerically. Their relation to the equations of DMF$^2$RG approach (or, more generally, the relation of 2PI to 1PI fRG approach) for the local bare interaction $V=\widetilde{V}=0$ is discussed in Appendix B. 


\subsection{Non-local susceptibilities and triangular vertices}
\label{NonlocSect}

The charge- and spin nonlocal susceptibilities can be obtained from the two-particle Green function (\ref{WVert}) as%
\begin{eqnarray}
\chi _{\Lambda ,q}^{c(s)} &=&-\sum\limits_{kk^{\prime }}\Pi _{\Lambda,k,q} \left( \delta _{kk^{\prime }}+{\mathcal F}_{\Lambda,kk^{\prime }q}^{c(s)}\Pi _{\Lambda,k^{\prime
},q} \right)  \nonumber \\
&=&-\sum\limits_{kk^{\prime }}\left[ \Pi _{\Lambda,k ,q }^{-1}\delta
_{kk'}+\widetilde \Phi _{\Lambda,k k' q
}^{(2),c(s)}-\Lambda (V_{\mathbf{q}}^{c(s)}-v^{c(s)}(i\omega _{n}))\right]
_{kk'}^{-1}.  \label{CDBL1}
\end{eqnarray}%
In the beginning of the flow the Green function $\overline{G}_{\Lambda=0 ,k}$ and
 susceptibilities $\chi_{\Lambda=0,q}^{c(s)}$ coincide with their local counterparts $G_{\rm loc}(i\nu_n)$ and $\chi_{\rm loc}^{c(s)}(i \omega_n)$, while in the end of the flow (at $\Lambda =1$) we obtain the non-local Green function $\overline{G}_{\Lambda=1 ,k%
 }=G_{0,k}^{-1}-\Sigma _{1,k}$ and non-local susceptibilities $\chi _{\Lambda=1 ,q}^{c(s)}$. 
 
 In the approximation, which neglects the flow of the two-particle irreducible vertices, the susceptibilities read
 \begin{eqnarray}
\chi _{\Lambda ,q}^{c(s)} &=&\int d\nu d\nu ^{\prime }\left[ (\chi _{\Lambda,\nu ,q }^{0})^{-1}\delta
_{\nu \nu ^{\prime }}-\Phi _{\mathrm{loc},\nu \nu ^{\prime }\omega
}^{(2),c(s)}+\Lambda (V_{\mathbf{q}}^{c(s)}-v^{c(s)}(i\omega _{n}))\right]
_{\nu \nu ^{\prime }}^{-1},  \label{CDBL}
\end{eqnarray}%
where 
\begin{equation}
\chi _{\Lambda ,\nu,q }^{0}=-\sum\limits_{\mathbf{k}}\Pi_{\Lambda,k,q}=-2\sum\limits_{\mathbf{k}}\overline{G}_{\Lambda ,%
\mathbf{k}\nu }\overline{G}_{\Lambda ,\mathbf{k+q},\nu +\omega },
\label{chi0}
\end{equation}%
and reproduce the result of the ladder
approximation in the dual boson approach (see, e.g., Refs. \cite{DB2,DB3}
and Appendix C) with fully renormalized Green's functions (including
non-local self-energy); for $v=0$ and $\Sigma_{1,k}=\Sigma_{\rm loc}(i \nu_n)$ (i.e. using DMFT as a starting
point and neglecting non-local self-energy corrections in the Green functions)  we also reproduce the result of ab initio ladder D$\Gamma $A approach \cite{abinitioDGA}. For $\Sigma_{1,k}=\Sigma_{\rm loc}(i \nu_n)$ the resulting susceptibility $\chi _{\Lambda=1 ,q}^{c}$ also fulfills charge conservation law, see Ref. \cite{DB3}.

The
susceptibilities (\ref{CDBL1}) in their general form can be also rewritten introducing triangular
vertices (cf. Ref. \cite{DGA1c})%
\begin{equation}
\gamma _{\Lambda,k ,q }^{c(s)}=\Pi _{\Lambda,k ,q }^{-1}\sum_{k'}\left[ \Pi _{\Lambda,k ,q }^{-1}\delta
_{kk'}{%
+\widetilde \Phi _{\Lambda,kk'q }^{(2),c(s)}\pm }U{/2%
}+v^{c(s)}(\omega)\right] ^{-1}
\label{gvert}
\end{equation}%
which yield random phase approximation (RPA)-like result (see Appendix C)%
\begin{equation}
\chi _{\Lambda ,q }^{c(s)}=\frac{\phi _{\Lambda ,q}^{c(s)}}{{1+}%
U_{\Lambda ,q}^{c(s)}\phi _{\Lambda ,q}^{c(s)}},  \label{chiL}
\end{equation}%
where 
\begin{equation}
U_{\Lambda ,q}^{c(s)}=\pm U/2+\Lambda V_{q}^{c(s)}+(1-\Lambda)v^{c(s)}(\omega )
\label{UL}
\end{equation}
is the corresponding $\Lambda$-dependent non-local interaction and 
\begin{equation}
\phi _{\Lambda ,q}^{c(s)}=
-\sum_k \gamma _{\Lambda,k ,q }^{c(s)}\Pi _{\Lambda,k ,q }  \label{phiL}
\end{equation}%
is the polarization operator (particle-hole irreducible susceptibility). The interaction (\ref{UL}) switches from fully local to non-local interaction when $\Lambda$ changes from zero to one, similarly as the equation (\ref{G0Lk}) switches from local to non-local Green function, such that the resulting flow describes the inclusion of non-local degrees of freedom in both, single-particle and two-particle interaction parts. The results (\ref{chiL}) and (\ref{phiL}) are similar to the susceptibilities in the ladder D$\Gamma$A approach \cite{DGA1c,abinitioDGA}, except that here we do not necessarily assume the locality of the 2PI vertices $\widetilde \Phi _{\Lambda,kk'q }^{(2),c(s)}$. Similarly to the $\lambda$-corrected D$\Gamma$A approach\cite{DGA1c,abinitioDGA} one can correct the susceptibility $(\chi^{s}_{\Lambda,q})^{-1}\rightarrow (\chi^{s}_{\Lambda,q})^{-1}+\lambda_\Lambda$, which also implies a correction of the 2PI vertex $\Phi^{s}_\Lambda\rightarrow\Phi^{s}_\Lambda-\lambda_\Lambda$, to 
fulfill a certain sum rule, e.g. $\sum_q \chi^s_{\Lambda,q}=\sum_\omega \chi^s_{\rm loc}(\omega)$. Applying the sum rule
avoids divergence of spin susceptibilities at low temperatures in two dimensions (and, therefore, allows to fulfill Mermin-Wagner theorem).

\subsection{The non-local correction to the self-energy}
\label{SESect}
The non-local corrections to self-energy can be obtained from the Eq.~(\ref%
{Sflow_eq}), which can be again rewritten in terms of the
triangular vertices (see diagrammatic form in Fig. \ref{FigDiagr}). By representing 
\begin{eqnarray}
{\Sigma }_{\Lambda ,k} &=&
2\Lambda\left[ V_{\mathbf{q=0}}^{c}-v^{c}(0)\right]
\sum\limits_{k^{\prime }}\overline{G}_{\Lambda ,k^{\prime }}  
-2\sum\limits_{k^{\prime }}\Phi _{\mathrm{loc},\nu \nu ^{\prime }0}^{(2),c}%
\left[ \overline{G}_{\Lambda ,k^{\prime }}- G_{\rm loc}(i\nu ^{\prime })%
\right] +{\hat{\Sigma }}_{\Lambda ,k},  \label{SF}
\end{eqnarray}%
with $\hat{\Sigma}_{\Lambda=0,k}=\Sigma_{\rm loc}(i \nu_n)$, we find%
\begin{eqnarray}
\frac{{d\hat{\Sigma }}_{\Lambda ,k}}{{d\Lambda }} &=&-2\sum%
\limits_{q,m=c,s}a_{m}^{(c)}\gamma _{\Lambda,k ,q }^{m}\frac{{\partial }%
{W}_{\Lambda ,q}^{m}}{{\partial \Lambda }}\gamma _{\Lambda ,k+q,-q}^{m}\overline{G}_{\Lambda ,k+q}  \nonumber \\
&&+2\sum\limits_{k^{\prime }}\left[ 
\Phi _{\mathrm{loc},\nu \nu ^{\prime }0}^{(2),c}
-\widetilde \Phi _{%
\Lambda,k k' 0}^{(2),c}\right] \frac{{d}\overline{G}%
_{\Lambda ,k^{\prime }}}{{d}\Lambda },  \label{dSigmaGW}
\end{eqnarray}%
where%
\begin{equation}
{W}_{\Lambda ,q}^{c(s)}=\frac{U_{\Lambda ,q}^{c(s)}}{{1+}U_{\Lambda
,q}^{c(s)}\phi _{\Lambda ,q}^{c(s)}}  \label{WL}
\end{equation}%
is the renormalized effective interaction and the partial $\Lambda $-derivative in the
right hand side of Eq. (\ref{dSigmaGW}) acts on $U_{\Lambda ,q}^{c(s)}$ only. The first term in the right-hand side of Eq. (\ref{SF}) represents change of the Hartree correction to the self-energy because of switch on the non-local interaction and can be absorbed into the chemical potential if we represent $\mu_\Lambda=\widetilde \mu _\Lambda+2\Lambda \left[ V_{\mathbf{q=0}}^{c}-v^{c}(0)\right]
\sum_{k^{\prime }}\overline{G}_{\Lambda ,k^{\prime }} $; according to the initial condition then $\widetilde \mu_{\Lambda=0}=\mu_{\rm loc}$. 
On the other hand, the second term in the Eq. (\ref{SF}) represents a correction to the local part of the self-energy due to change of the local Green function with $\Lambda$. We do not consider this correction in the following, since we assume that (E)DMFT provides a correct local starting point (as shown e.g., by the comparison to numerical results for the self-energy in Sect. \ref{NumericSigma}).
We therefore consider 
$\hat{\Sigma}_{\Lambda=1}$ as the physical 
self-energy up to the above mentioned shift of the chemical potential. 


The equation (\ref{dSigmaGW})
has a differential form, which differs the considered approach from previously considered non-local extentions of EDMFT. The first term in this equation has the structure, which is similar to the
DB-GW$\gamma$ \cite{DB4}, and TRILEX \cite{TRILEX} approaches; 
in contrast to these approaches it however describes mainly the contribution of the non-local interaction (note that ${\partial }{W}_{\Lambda ,q}^{m}/{\partial \Lambda }=0$ in the absence of non-local interaction). The contribution of the local interaction is accounted mainly via the second term in Eq. (\ref{dSigmaGW}), although the contributions of both types of interactions are mixed because of the renormalization of the vertices.
Despite the similarity to EDMFT+GW approach, the considered approach essentially improves the results of the former method (see next Section) due to account of non-local four- and three-point (triangular) vertices in Eqs. (\ref{phiL}) and (\ref{dSigmaGW}). 

The approximation, which keeps only the ladder diagrams in the non-local self-energy (\ref{dSigmaGW}) (denoted in the following as the ladder approximation)
can be obtained by using the transverse-channel ladder approximation (\ref{Phi21st}) in the second term of the right-hand side of Eq. (\ref{dSigmaGW}). At the same time, in the first term of this equation, as well as in the corresponding susceptibilities  (\ref{CDBL}) and triangular vertices (\ref{gvert}), it is consistent then to use $\widetilde \Phi _{\Lambda,k k' q
}^{(2),c(s)}=\widetilde \Phi _{\Lambda=0,k k' q
}^{(2),c(s)}=\Phi _{{\rm loc},\nu \nu' \omega
}^{(2),c(s)}$ to stay on the level of ladder diagrams. 
The results of the ladder and non-ladder approximations are considered in the next Section.

Note that in general, the 2PI vertices may suffer from the divergences \cite{Toschi,div1,div2}, which are likely related to the property of Luttinger-Ward functional being not uniquely defined \cite{BKFail,BKFail1}. Although it is not obvious how these divergences can be circumvented in general case (which we postpone to future studies), we stress that the 1PI vertices $\mathcal F^m$ in the right-hand side of Eqs. (\ref{flow_eq}) are well defined (and not divergent) for the bare (E)DMFT 2PI vertices. In particular, the above discussed ladder approximation, which contains only the vertices $\mathcal F^m$ with the bare local 2PI vertices (in view of Eqs. (\ref{Phi21st}), (\ref{dSigmaGW})), does not suffer from the mentioned divergences.  
This also allows us to suppose that if the 2PI vertices do not change strongly with respect to their bare values, one can expect that the right-hand sides of Eqs. (\ref{ifflow_eq}) and (\ref{dSigmaGW}) are still well defined. We show below that at least sufficiently close to the divergence the equations (\ref{flow_eq}) in a certain vertex projection scheme do not lose their applicability.
\section{Numerical implementation and results}

For numerical implementation of (E)DMFT we use
hybridization expansion continous-time QMC method within iQIST package of Refs. \cite{iQIST,iQIST1},
choosing $N_{b}=46$ to $76$ non-negative bosonic and $N_{f}=60$ to $180$
fermionic Matsubara frequencies  for the vertex calculation. We use the 2PI vertices obtained in (E)DMFT (without performing further adjustment of bath Green function) as an input of Eqs.~(\ref{Phi21st}) and (\ref{dSigmaGW}) in the ladder approach and Eqs.~(\ref{ifflow_eq}) and (\ref{dSigmaGW}) in the non-ladder approach, and account for the symmetries of the 
two fermion and fermion-boson 
vertices considered in Ref. \cite{DB2}.

In the numerical implementation of the 2PI fRG approach we consider the truncation of Eq. (\ref{ifflow_eq}), which is restricted to the contribution of charge and spin vertices in the right-hand side. In the non-ladder 2PI fRG approach we parametrize the momentum dependence of the corresponding irreducible vertices as $\widetilde \Phi^{c(s)}_{\Lambda,kk'q}=\varphi^{c(s)}_{\Lambda,\nu\nu'\omega}({\bf k'-k})$, i.e. assume that they depend strongly on the momentum transfer $\bf k'-k$  only. This is motivated by the form of the right-hand side of Eq. (\ref{ifflow_eq}), as well as the ladder approximation (\ref{Phi21st}). To simplify solution of Bethe-Salpeter equations, which determine $\mathcal F^{c(s)}$ in the right-hand side of Eqs. (\ref{flow_eq}) and (\ref{dSigmaGW}), we approximate the vertices $\widetilde \Phi^{c(s)}$ in the Bethe-Salpeter equation (\ref{GammaVert1}) by their local values $\sum_{\bf p} \varphi^{c(s)}_{\Lambda,\nu\nu'\omega}({\bf p})$. This approximation can be considered as the lowest order approximation projecting momentum dependences of $\varphi_\Lambda^m(\bf p)$ onto set of the form factors, among which we choose the constant form factor only, cf. Refs. \cite{Salmhofer,DMF2RG3}. For parameterization of momentum dependences we choose 10$\times$10 momenta points in each quadrant of the Brillouin zone. The maximal numerical effort for the solution of fRG equations in this form is approximately 500 core$*$hours, which is an order of magnitude smaller than calculating (E)DMFT vertices for considered number of frequencies.

\subsection{Numerical results for the local bare interaction}\label{NumericSigma}

\begin{figure}[t]
\center \includegraphics[width=0.7\linewidth]{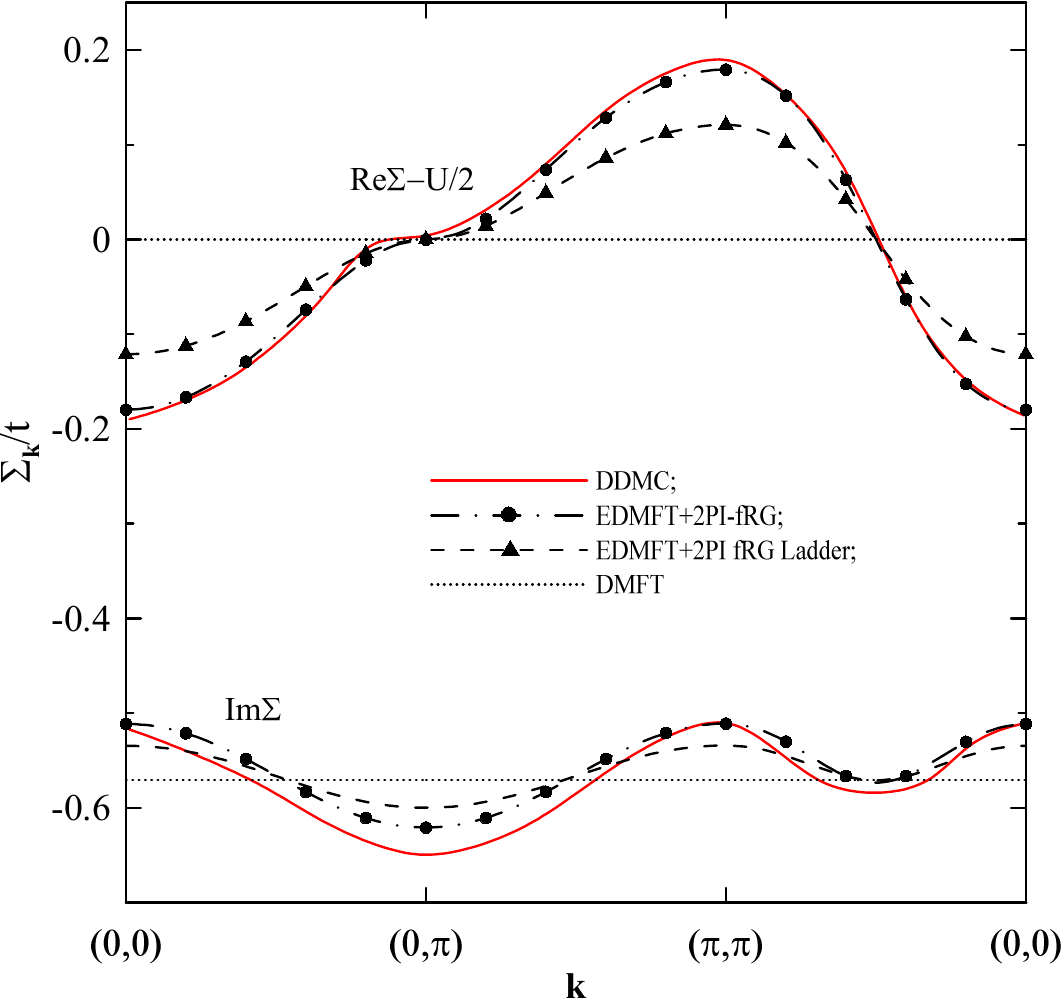} \endcenter
\caption{(Color online) Real (upper set of lines) and imaginary (lower set of lines) parts of $\Sigma_{\Lambda=1,{\bf k},\nu_1}$ at first Matsubara frequency $\nu_1=\pi T$ for Hubbard model with $U=4t$, $T=0.5t$ in the EDMFT+2PI-fRG approach (dot-dashed lines, circles), as well as the ladder EDMFT+2PI-fRG approach (dashed lines, triangles). Symbols mark each second point of the momentum grid in selected directions. For comparison, the DDMC results of Ref. \cite{BKFail} (solid lines) are presented.}
\label{FigSigma}
\end{figure}

As an example of the application of the developed approach, we consider first the two dimensional half filled
Hubbard model on the square lattice with $\epsilon_{\mathbf{k}}=-2t(\cos
k_{x}+\cos k_{y})$, $\widetilde \mu_\Lambda=\mu_{\rm loc}$, $V^{c(s)}_{\bf q}=v^{c(s)}(i\omega)=0$. We choose the interaction $U=4t$, which is considered sufficiently close to the divergence of charge 2PI vertex \cite{Toschi,BKFail}. Indeed, the DMFT calculation yields $\Phi^{(2),c}_{{\rm loc},\nu_1,\nu_1,0}\simeq -12t$ at the considered temperature $T=0.5t$, where $\nu_1=\pi T$ is the first fermionic Matsubara frequency. In Fig. \ref{FigSigma} we present the results for the self-energy and compare them to the results of diagrammatic determinant Monte Carlo (DDMC) method \cite{BKFail}, which are also close to the results of Blankenbecler-Sugar-Scalapino quantum Monte Carlo approach \cite{BSS-QMC}.
One can see that the considered truncation yields almost correct real part of the self-energy and slightly underestimates non-local contribution to the imaginary part. We also compare the obtained results to the results of the solution of Eq. (\ref{dSigmaGW}) 
in the ladder approximation (\ref{Phi21st}). The ladder approximation yields smaller non-local correction to the self-energy, yet it reproduces qualitatively correct the momentum dependence of the self-energy.

\subsection{Application to the $U$-$V$ Hubbard model}

Let us consider next the application of the developed method to studying charge instability in the two dimensional extended $U$-$V$
half filled Hubbard model on the square lattice with 
$V_{\mathbf{q}}^{c}=2V(\cos
q_{x}+\cos q_{y})$, $V_{\mathbf{q}}^{s}=0$, and the same dispersion $\epsilon_{\mathbf{k}}$ as in Sect. \ref{NumericSigma},
which was chosen as a test for
previously developed approaches\cite{DB2,DB4,Ayral1,AyralGW,Ayral,DCA}. As a starting point of
2PI-fRG scheme we choose EDMFT solution with $v^{s}(\omega )=0$ and $v^{c}(\omega ),
$ fulfilling Eq. (\ref{sc1}). 
We then solve fRG equations (\ref{ifflow_eq}) and (\ref{dSigmaGW}) with the parameterization of vertices,  described in the beginning of this Section.
To study sufficiently low temperatures we introduce $\lambda_\Lambda$-correction to the vertex $\Phi^{s}_\Lambda$ as described in Sect. \ref{NonlocSect}. 

We detect charge instability by vanishing inverse charge susceptibility (\ref{chiL}) in the end of the flow. Since the charge density wave susceptibility with the wave vector ${\mathbf Q}=(\pi,\pi)$ diverges most strongly in the considering case, we consider only this instability; the condition for the instability has mean-field-like or RPA-like form
\begin{equation}
{1+}(U/2- 4V)\phi _{\Lambda=1 ,{\mathbf Q},\omega=0}^{c}=0.  
\label{PhaseBoundary}
\end{equation}
The renormalized polarization operator $\phi^c$ contains, however, in contrast to RPA, self-energy and vertex corrections according to the Eqs. (\ref{gvert}) and (\ref{phiL}). 

\begin{figure}[tbp]
\center 
\vspace{.5cm}
\includegraphics[width=0.7\linewidth]{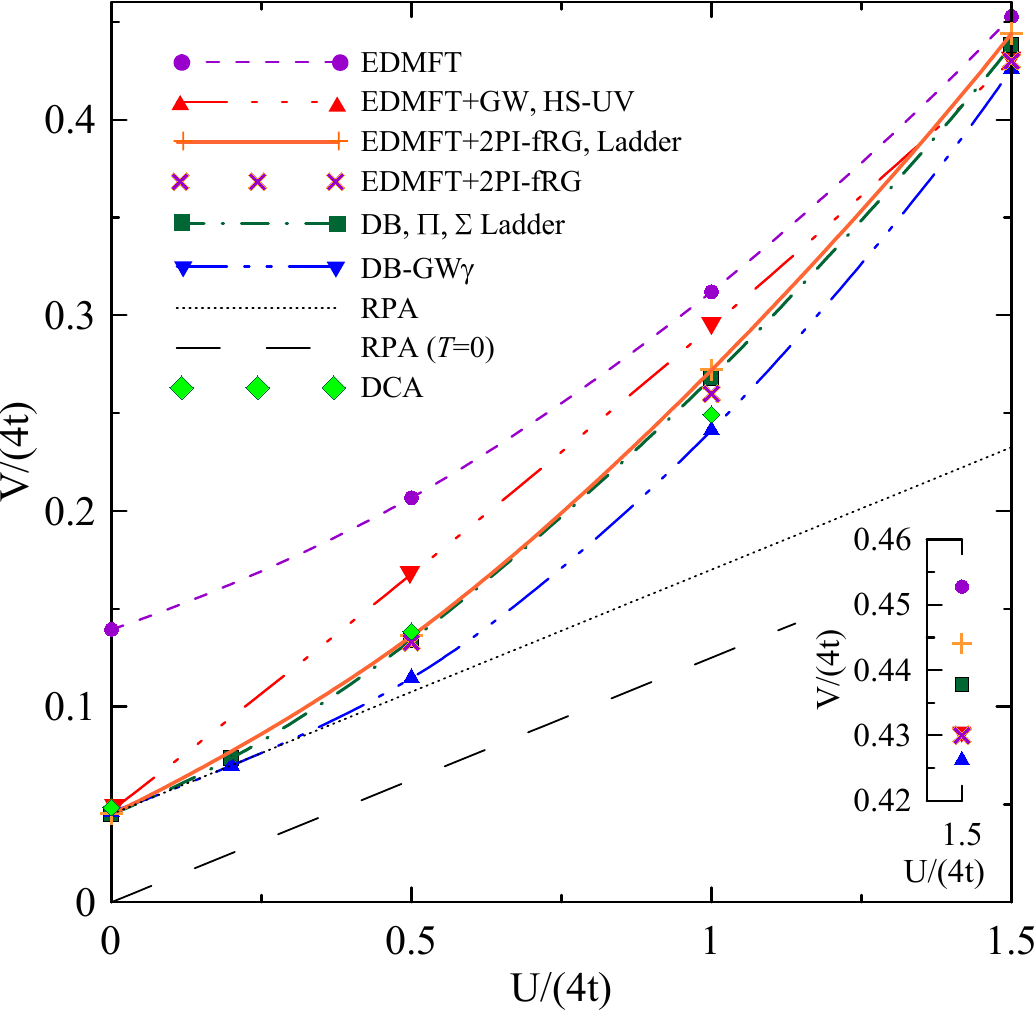} 
\endcenter
\caption{(Color online) Phase boundary of charge density wave instability at $\beta=1/T=12.5t^{-1}$, obtained in the EDMFT+2PI-fRG ladder (solid orange line, crosses) and the non-ladder approach (violet crosses), compared to the results of the EDMFT approach \cite{Ayral1,DB2} (dashed violet line with circles), dual boson (DB, $\Pi$, $\Sigma$) ladder approach   \cite{DB2} (dot-dashed green line, squares), the DB-GW$\gamma$ approach \cite{DB4} (dot-dot-dashed blue line, triangles), EDMFT+GW approach in HS-UV decoupling (red dot-dot-dot dashed line, downward triangles) and dynamic cluster approximation \cite{DCA} (green rhombs). Dotted line denotes the result of RPA and long-dashed black line corresponds to RPA at $T=0$, $V^c_{{\rm RPA},T=0}=U/8$. The lower-right inset shows zoom of the results for $U=6t$.}
\label{PhaseD}
\end{figure}

The boundaries of the charge density wave instability in the presented approaches at the temperature $T=0.08t$ are shown and compared to earlier results in Fig. \ref{PhaseD}. 
\begin{figure}[t]
\center 
\includegraphics[width=0.7\linewidth]{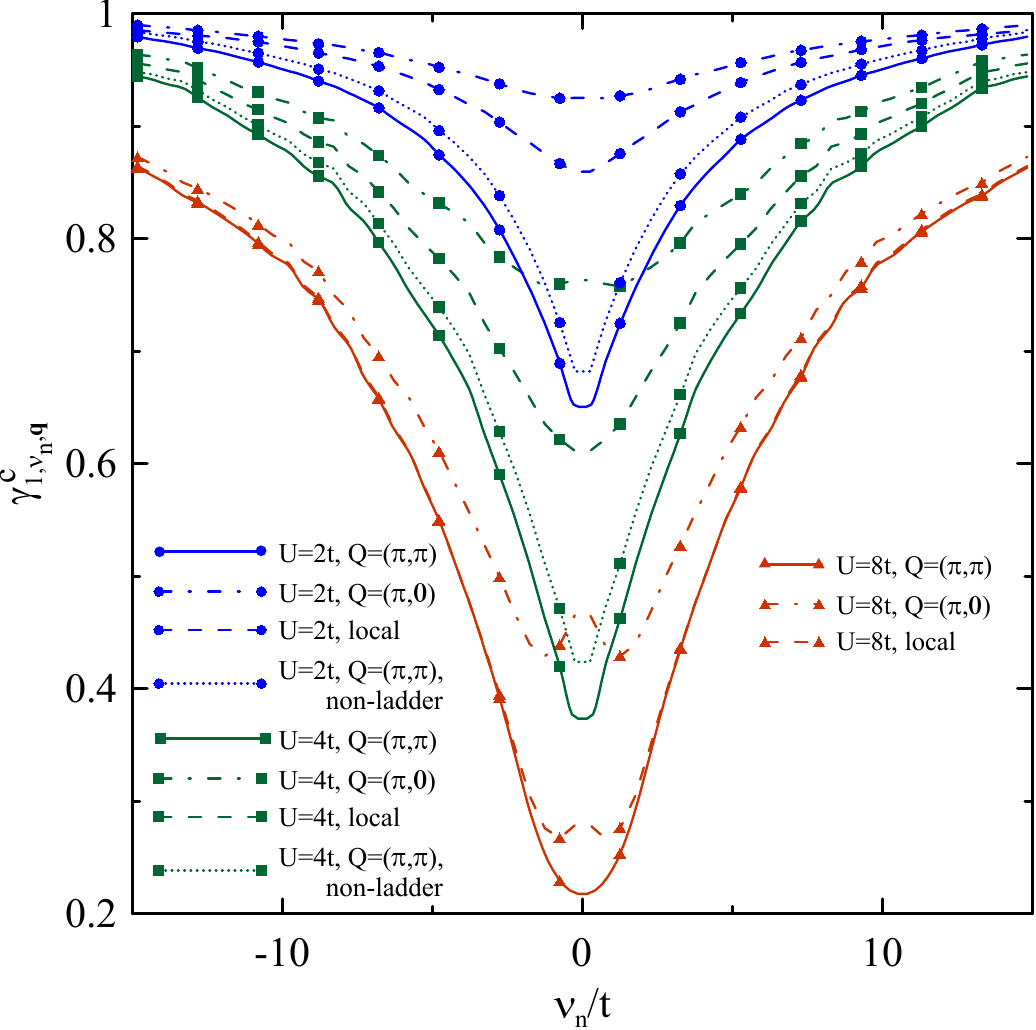}
\endcenter
\caption{(Color online) Fermionic frequency dependence of static triangular vertices $\gamma_{1,\nu_n,\mathbf{q}}$ in the charge channel at $\beta=1/T=12.5t^{-1}$, obtained in the end of the flow of the ladder EDMFT+2PI-fRG approach for $\mathbf{q}=(\pi,0)$ (dash-doted lines), $\mathbf{q}=(\pi,\pi)$ (solid lines), compared to the local vertices of the EDMFT approach (dashed line) for $U=2t$, $V=0.524t$ (blue lines, circles), $U=4t$, $V=1.02t$ (green lines, squares), and $U=8t$, $V=2.592t$ (red lines, triangles) near the phase boundary to CDW instability. Dotted lines show the triangular vertex at $\mathbf{q}=(\pi,\pi)$ in EDMFT+2PI-fRG approach beyond ladder approximation (see text). Symbols mark each fourth Matsubara frequency.}
\label{GammaVert}
\end{figure}
The results of the ladder approximation, discussed in the end of Sect. \ref{SESect} are close to the results, obtained within DB approach \cite{DB2}. Due to account of non-local corrections to triangular vertices (\ref{gvert}), the considered methods yield better agreement with the DB approach, than the DB-GW$\gamma$ approach of Ref. \cite{DB4}, which uses local triangular vertices. The obtained momentum dependence of the triangular vertices is rather strong and shown in Fig. \ref{GammaVert}. At sufficiently large interactions $U$ we find that the renormalized polarization operator $\phi_{\Lambda,{\mathbf q},\omega=0}$ becomes weakly momentum dependent, and, therefore, approaches its local value. 
While this weak momentum dependence of $\phi_{\Lambda,{\mathbf q},\omega=0}$ appears as a result of peculiar frequency dependence of the particle-hole irreducible vertex $\Phi^{(2),c}_{\rm loc,\nu\nu'0}$, it can be also viewed as a cancellation of momentum dependence of the bare susceptibility $\chi^0_{\Lambda,\nu,q}$ and the vertex $\gamma^c_{\Lambda,\nu,q}$ in Eq. (\ref{phiL}). We have verified that slight increase of the obtained critical $V_c$ of charge instability in the ladder approximation in comparison to the results of the dual boson approach is mainly due to the effect, which was not accounted in Ref. \cite{DB2}, namely the contribution of non-local spin correlations (the terms, containing $\mathcal F^s_\Lambda$ in the right-hand side of Eq. (\ref{Phi21st})) to the self-energy (\ref{dSigmaGW}).  
Although both, ladder and non-ladder EDMFT+2PI-fRG approaches agree well with recent dynamic cluster approximation (DCA) study \cite{DCA}, interestingly enough, the fRG analysis within Eq. (\ref{ifflow_eq}) beyond the ladder approximation (which we have performed up to $U=6t$) yields somewhat smaller critical interaction and better agreement with DCA data of Ref. \cite{DCA} for $U=4t$. This can be attributed to another effect, not contained in ladder versions of DB and EDMFT+2PI-fRG approaches, namely to the renormalization of 2PI charge vertex by charge and (mainly) spin correlations in the ``transverse" channel, which enhance charge instability, since $a^{(c)}_{c,s}>0$, and countervail the above mentioned effect of increase of critical interaction because of the increase of electronic damping due to spin correlations. 

We note also that EDMFT+GW approach in HS-$V$ decoupling form \cite{AyralGW} yields much smaller critical interaction $V$ than the above discussed approaches. The results of HS-$UV$ decoupling of EDMFT+GW approach, obtained in Ref. \cite{Ayral}, are shown in Fig. \ref{PhaseD} and they are numerically closer to the dual boson approach, but the slope of the boundary of charge instability at small $V$ is strongly different from the above discussed approaches.

The boundary of the stability of charge density wave phase, obtained in mean-field (or RPA) approach follows from Eq. (\ref{PhaseBoundary}) by replacing dressed polarization operator $\phi^c$ with the bare one, $\phi^{c,0}_q=-2\sum_k G_{0,k}G_{0,k+q}$ and can be written in the form $V^c_{\rm RPA}=U/8+1/(4\phi^{c,0}_{{\bf Q},\omega=0})$  (see also Ref. \cite{Note}). One can see that it has the same slope at small $U$ as the obtained phase boundary in more sophisticated approaches, but strongly underestimates the obtained critical interaction $V$ at finite $U$. While this holds at finite temperatures considered, it is interesting to discuss the possibility of charge instability at $T\rightarrow 0$. In the weak-to-intermediate coupling regime  
assuming Fermi-liguid form of the fermionic self-energy (which according to our results holds for the studied interaction $U$ range), one expects that  $\phi _{\Lambda=1 ,{\mathbf Q},\omega=0}$ diverges logarithmically at $T\rightarrow 0$, and therefore 
charge instability occurs at $V^c_{{\rm RPA},T=0}=U/8$ in both, RPA and the considered approach, including self-energy and vertex corrections (the same result applies to the DB approach as well). This logarithmic divergence is, however, 
weakened by the vertex corrections,
since, according to the Eq. (\ref{phiL}) we find  $\phi _{\Lambda=1 ,{\mathbf Q},\omega=0}\propto \gamma^c _{\Lambda=1 ,\nu\rightarrow 0,{\mathbf Q},\omega=0} \ln(t/T)$ and the vertices $\gamma^c$ are suppressed by both, local and non-local correlations (see Fig. \ref{GammaVert}). This implies 
suppression of transition temperatures at intermediate 
$U$ due to above mentioned vertex corrections. On the other hand, as mentioned above, in the strong coupling regime the renormalized polarization operator $\phi _{\Lambda=1 ,{\mathbf Q},\omega=0}$ approaches its local value, and, therefore it is weakly temperature dependent. In this regime even at $T\rightarrow 0$ the boundary of charge instability is expected to approach EDMFT result. The crossover or transition between the two regimes (corresponding to change from itinerant to localized behavior) will be studied elsewhere.
Also, 
at $T\rightarrow0$ charge instability competes with the spin density wave and therefore the former may become dominant instability in the ground state at larger value of the non-local interaction $V^{\rm c,dom}$ than the interaction $V^c$ determined from the vanishing of inverse charge susceptibility (e.g., in the RPA comparison of the inverse charge and spin susceptibilities yields $V^{\rm c,dom}_{{\rm RPA},T=0}=U/4$).

\section{Conclusions}

In conclusion, we have presented a general EDMFT+2PI-fRG approach, which
considers the 2PI functional renormalization-group flow, starting from the
(extended) dynamical mean-field theory. 
The considered approach operates directly with the physical interaction ${W}_{\Lambda ,q}$, whose $\Lambda$-dependence reflects only to which extent the bare local Coulomb interaction $v(\omega)$ is replaced by its non-local counterpart $V_{\mathbf q}$, and the physical Green functions $\overline{G}_{\Lambda ,k}$, where also $\Lambda$-dependence reflects growing effect of the non-local contributions (replacing local bath Green function of (E)DMFT by the non-local lattice one and introducing the non-local self-energy part) instead of the corresponding dual quantities. 

We have
shown that for purely local interactions the considering approach describes non-local corrections to the self-energy due to charge- and spin correlations. 
For
the non-local interaction, in the simplest truncation of scale-independent 2PI vertices the susceptibilities in the considering approach have the same form as in the
ladder approximation in the dual boson (DB) \cite{DB1,DB2,DB3} and ab initio
D$\Gamma $A approach \cite{abinitioDGA}. At the same time, the EDMFT+2PI-fRG approach allows to consider an interplay of charge- and spin correlations in the presence of non-local interactions.  


We have tested the proposed approach
by comparing the self-energy for purely local interaction with the results of numerical DDMC calculations and by studying the possibility of charge instability in the half filled two-dimensional extended $U-V$ model. For the latter model we have shown that the considered method allows to obtain results, which are close to the dual boson approach and dynamic cluster approximation, improving the dual boson approach by treatment of the effect of spin correlations on charge instability. 
We have traced the origin of strong enhancement of critical interaction $V^c$ in comparison to the mean-field (RPA) result $V^c_{\rm RPA}=U/8+{\rm const}$ in the intermediate-to-strong coupling regime, which appears because of substantial local and non-local vertex corrections. We have also shown that the effect of spin correlations on charge density wave phase boundary is small, and leads to weak decrease of critical next-nearest neighbor repulsion for charge instability in comparison to the ladder approximation. 

Although the current version of the approach is in general not conserving, since the used truncations (neglecting Aslamazov-Larkin diagrams, approximation of $\Phi^{(3)}$ vertex, expressing it through the two-particle vertices, and neglecting contributions of higher order vertices) violate conservation laws, improvements of the proposed approach accounting for the Aslamazov-Larkin diagrams and using better approximations for the vertices $\Phi^{(3,4)}$ are expected to provide better fulfillment of conservation laws and should be investigated in future.

Similarly to the dual fermion and dual boson approaches, the presented approach can be also applied self-consistently: the non-local
self-energy, determined in the end of the flow, can produce the new local
Green function, which allows to adjust the bath Green functions of the local problem. Investigation of this possibility is also postponed for future studies. 

The proposed scheme is rather general, and more sophisticated truncations
can be used to improve the results of the mentioned approaches. Numerical
investigations of the presented equations will allow to study the concrete
phenomena, such as charge- or spin-density wave instabilities in
strongly-correlated systems, as well as screening of the long-range Coulomb
interaction in the presence of strong electronic correlations.

\textit{Acknowledgements. }The author is grateful to A. I. Lichtenstein, E.~v.~Loon, A.~Toschi, and C. Taranto for
stimulating discussions. The work is performed within the theme ``Quant"
AAAA-A18-118020190095-4 of FASO, Russian Federation. The calculations are performed on the ``Uran" cluster of UB RAS. 

\appendix
\section{Derivation of 2PI fRG equations for self-energy and
vertex}

To derive the 2PI-fRG equations, we follow the standard strategy, outlined
in Refs. \cite{Dupuis,Meden}. Differentiating Eqs. (\ref{ZJ}) and (\ref%
{GammaG}) with respect to $\Lambda $, we find ($W_{\Lambda }=\ln Z_{\Lambda
} $): 
\begin{eqnarray}
\dot{W}_{\Lambda }[J] &=&\frac{1}{Z_\Lambda}\sum_{k}\frac{\partial G_{0\Lambda
,k}^{-1}}{\partial \Lambda }\frac{\delta Z_{\Lambda }}{\delta J_{k,0}^c}-%
\frac{1}{2Z_\Lambda}\sum_{kk^{\prime }q,m}\widetilde{V}_{q}^{m}\frac{\delta
^{2}Z_{\Lambda }}{\delta J_{kq}^m \delta J_{k^{\prime },-q}^m }  \nonumber \\
&=&-\frac{1}{2}\sum_{kk^{\prime }q,m}\widetilde{V}_{q}^{m}\left( \frac{%
\delta ^{2}W_{\Lambda }}{\delta J_{kq}^m \delta J_{k^{\prime },-q}^m }+\frac{%
\delta W_{\Lambda }}{\delta J_{kq}^m}\frac{\delta W_{\Lambda }}{\delta
J_{k^{\prime },-q}^m }\right)  \nonumber \\
&+&\sum_{k}\frac{\partial G_{0\Lambda ,k}^{-1}}{\partial \Lambda }%
\frac{\delta W_{\Lambda }}{\delta J_{k,0}^c }.
\end{eqnarray}
The first derivative of $W_\Lambda$ is obtained from Eq. (\ref{Gamma_first}) as $\delta W_\Lambda/\delta J^m_{k,q}=G^m_{k,k+q}$, while for obtaining second derivative we differentiate Eq. (\ref{GammaG}) twice:
\begin{eqnarray}
\Gamma _{\Lambda ,kk^{\prime }q}^{(2),mm^{\prime }} &\equiv &\frac{\delta
^{2}\Gamma_\Lambda }{\delta G_{k,k+q}^m \delta G_{k^{\prime }+q,k'}^{m^{\prime }}}=-[\Pi _{kk^{\prime }q}^{mm^{\prime }}]^{-1}-\Phi _{\Lambda ,kk^{\prime
}q}^{(2),mm^{\prime }}  
=\left[ W_{\Lambda ,kk^{\prime }q}^{(2),mm^{\prime }}\right]
^{-1},
\end{eqnarray}
where $[\Pi _{kk^{\prime }q}^{mm^{\prime }}]^{-1}$
is defined after Eq. (\ref{PhiRG}) and $%
W_{\Lambda ,kk^{\prime }q}^{(2),mm^{\prime }}=\left[ \Gamma _{\Lambda
,kk^{\prime }q}^{(2),mm^{\prime }}
\right]
^{-1}=-\left[ \Pi ^{-1}+\Phi _{\Lambda }^{(2)}
\right]
_{kk^{\prime }q,mm^{\prime }}^{-1}$ is the corresponding susceptibility, the
inversion is performed with respect to momentum $k,k^{\prime }$ and channel $%
m,m^{\prime }$ indices. 
From this we find
\begin{eqnarray}
\dot{\Gamma }_{\Lambda }[G] &=&-\frac{1}{2}\sum_{kk^{\prime }q,m}\widetilde{V%
}_{q}^{m}\left[ \Pi ^{-1}+\Phi _{\Lambda }^{(2)}
\right]
_{kk^{\prime }q,mm}^{-1}-\sum_{k}\frac{\partial G_{0\Lambda
,k}^{-1}}{\partial \Lambda }G_{k,k}^{c}   \\
&&+\frac{1}{2}\sum_{q,kk^{\prime },m
}\left. \widetilde{V}%
_{q}^{m}
\right.
G_{k,k+q}^m G_{k^{\prime }+q,k'}^{m}, \nonumber \\
\dot{\Phi }_{\Lambda }[G] &=&\frac{1}{2}\sum_{kk^{\prime }q,m}\widetilde{V}%
_{q}^{m}\left[ \Pi ^{-1}+\Phi _{\Lambda }^{(2)}
\right]
_{kk^{\prime }q,mm}^{-1}  -\frac{1}{2}\sum_{kk^{\prime }q,m
}\left. \widetilde{V}%
_{q}^{m}
\right.
G_{k,k+q}^m G_{k^{\prime }+q,k'}^{m}.\label{Phidot}
\end{eqnarray}\vspace{-0.1cm}%
Taking variational derivatives over $G$ we obtain%
\begin{eqnarray}
\dot{\Phi }_{\Lambda ,k,q}^{(1),m} &=&\frac{\delta \dot{\Phi }_{\Lambda }}{%
\delta G_{k,k+q}^m }=\frac{1}{2}\sum_{k^{\prime }k^{\prime \prime }q^{\prime
},m^{\prime }}\widetilde{V}_{q^{\prime }}^{m^{\prime }}\left\{ \left[ \Pi
^{-1}+\Phi _{\Lambda }^{(2)}
\right] ^{-1}\right.\Pi ^{-1}  \label{A5} \\
&&\left. \times \frac{\delta \Pi }{\delta G_{k,k+q}^{m}}\Pi ^{-1}\left[
\Pi ^{-1}+\Phi _{\Lambda }^{(2)}
\right] ^{-1}\right\}
_{k^{\prime }k^{\prime \prime }q^{\prime },m^{\prime }m^{\prime }} -\widetilde{V}_{q}^{m}\sum_{k^{\prime }}G_{k^{\prime
}+q,k'}^{m} \nonumber
\end{eqnarray}
\begin{eqnarray}
\dot{\Phi }_{\Lambda ,kk^{\prime }q}^{(2),mm^{\prime }} &=&\frac{\delta \dot{%
\Phi }_{\Lambda }}{\delta G_{k,k+q}^{m}\delta G_{k^{\prime }+q,k'}^{m^{\prime }}}=%
-\left.\widetilde{V}_{q}^{m}\delta _{mm^{\prime }}
\right. \nonumber \\
&&+\frac{1}{2}\sum_{k^{\prime \prime }k^{\prime \prime \prime }q^{\prime
}m^{\prime \prime }}\widetilde{V}_{q^{\prime }}^{m^{\prime \prime }}\left\{ %
\left[ \Pi ^{-1}+\Phi _{\Lambda }^{(2)}
\right] ^{-1}\Pi ^{-1}%
\frac{\delta \Pi }{\delta G_{k^{\prime }+q,k'}^{m^{\prime }}}\Pi ^{-1}\right. 
\nonumber \\
&&\times \left[ \Pi ^{-1}+\Phi _{\Lambda }^{(2)}
\right]
^{-1}\Pi ^{-1}\frac{\delta \Pi }{\delta G_{k,k+q}^{m}}\Pi ^{-1}\left[ \Pi
^{-1}+\Phi _{\Lambda }^{(2)}
\right] ^{-1}  \nonumber \\
&&-\widetilde{V}_{q^{\prime }}^{m^{\prime \prime }}\left[ \Pi ^{-1}+\Phi
_{\Lambda }^{(2)}
\right] ^{-1}\Pi ^{-1}\frac{\delta \Pi }{%
\delta G_{k,k+q}^{m}}\Pi ^{-1}\frac{\delta \Pi }{\delta G_{k^{\prime
}+q,k'}^{m^{\prime }}} \Pi ^{-1}  \nonumber \\
&&\left. \times\left[ \Pi ^{-1}+\Phi _{\Lambda }^{(2)}
\right]
^{-1}+(k%
\begin{array}{c}
\leftrightarrow%
\end{array}%
k^{\prime },q%
\begin{array}{c}
\leftrightarrow%
\end{array}%
-q,m%
\begin{array}{c}
\leftrightarrow%
\end{array}%
m^{\prime })\right\} _{k^{\prime \prime }k^{\prime \prime \prime }q^{\prime
},m^{\prime \prime }m^{\prime \prime }}  \nonumber \\
&&-\frac{1}{2}\sum_{k^{\prime \prime }k^{\prime \prime \prime }m^{\prime
\prime }}\widetilde{V}_{k^{\prime }-k}^{m^{\prime \prime }}\left\{ \left[
\Pi ^{-1}+\Phi _{\Lambda }^{(2)}
\right] ^{-1}\Pi ^{-1}\frac{%
\delta ^{2}\Pi }{\delta G_{k,k+q}^{m}\delta G_{k^{\prime }+q,k'}^{m^{\prime }}}\Pi
^{-1}\right.  \nonumber \\
&&\times \left. \left[ \Pi ^{-1}+\Phi _{\Lambda }^{(2)}
\right]
^{-1}\right\} _{k^{\prime \prime }k^{\prime \prime \prime },k^{\prime
}-k,m^{\prime \prime }m^{\prime \prime }}.\label{A6}
\end{eqnarray}%
In the derivation of equations (\ref{A5}) and (\ref{A6}) 
we have neglected the dependence of $\Phi^{(2)}$ in the right-hand side of Eq.  (\ref{Phidot}) on $G$, which implies neglecting contributions, containing higher-order vertices ($\Phi^{(3)}$ for $\dot{\Phi }^{(1)}$ and $\Phi^{(3,4)}$ for $\dot{\Phi }^{(2)}$), cf. Refs. \cite{Dupuis,Meden}. This corresponds to neglect of the contribution of three- and four- particle processes to the renormalization of one- and two-particle vertices, keeping only contribution of the two-particle processes. While the neglected contributions may be important to describe critical behavior near phase transitions (e.g., four-particle contributions correspond in terms of bosonic degrees of freedom $\phi$ to the $\phi^4$ interaction between critical modes), their consideration is beyond the scope of the present paper.

An explicit calculation of the polarization operators yields%
\begin{eqnarray}
\Pi _{kk^{\prime }q}^{c(s),c(s)} &=&\left( G_{k,k^{\prime }}^{c}G_{k^{\prime
}+q,k+q}^{c}+G_{k,k^{\prime }}^{s}G_{k^{\prime }+q,k+q}^{s}\pm G_{k,k^{\prime }}^{+}G_{k^{\prime }+q,k+q}^{-}\right) /2, 
\nonumber \\
\Pi _{kk^{\prime }q}^{+-} &=&G_{k,k^{\prime }}^{c}G_{k^{\prime }+q,k+q}^{c}-G_{k,k^{\prime }}^{s}G_{k^{\prime }+q,k+q}^{s},  \nonumber \\
\Pi _{kk^{\prime }q}^{cs} &=&(G_{k,k^{\prime }}^{c}G_{k^{\prime
}+q,k+q}^{s}+G_{k,k^{\prime }}^{s}G_{k^{\prime }+q,k+q}^{c})/2,\ \   \nonumber \\
\Pi _{kk^{\prime }q}^{c(s),\pm } &=&(G_{k,k^{\prime }}^{c(s)}G_{k^{\prime
}+q,k+q}^{\pm }+G_{k,k^{\prime }}^{\pm }G_{k^{\prime }+q,k+q}^{c(s)})/2,
\end{eqnarray}%
where $G_{kk'}^{s}$ corresponds to $s_{z}$ component of the Green function in
the spin channel, and $G_{kk'}^{\pm }$ corresponds to its $s_{x}\pm i s_y$ components.
 Simplifying, we find at the stationary point%

\begin{eqnarray}
\dot{\overline\Phi }_{\Lambda,k}^{(1),m} &\equiv &\dot{\overline\Phi }_{\Lambda,k,0}^{(1),m}=\sum_{q,m^{%
\prime }=c,s}a_{m^{\prime }}^{(m)}\left[\overline \Pi_\Lambda ^{-1}\overline W_{\Lambda }^{(2)}\widetilde{V}%
_{q}^{m^{\prime }}\overline W_{\Lambda }^{(2)}\overline \Pi_\Lambda ^{-1}\right] _{k,k,q,m^{\prime
},m^{\prime }}{\overline G}_{\Lambda,k+q}^{m}  \nonumber \\
&&-\widetilde{V}_{q=0}^{m}\sum_{k^{\prime }}
{\overline G}_{\Lambda,k^{\prime
}}^{m}+\sum\limits_{k^{\prime },m'}\overline{\Phi }_{\Lambda,kk^{\prime }0}^{(2),mm'}%
\frac{{d}\overline{G}^{m'}_{\Lambda ,k^{\prime }}}{{d}\Lambda }, \label{G1} \\
\dot{\overline\Phi }_{\Lambda,kk^{\prime }q}^{(2),mm^{\prime }} &=&\sum_{q^{\prime
}m^{\prime \prime }=c,s}c_{m^{\prime \prime }m^{\prime \prime \prime
}}^{(m,m^{\prime })}\left\{ [\overline\Pi_\Lambda ^{-1}\overline W_{\Lambda }^{(2)}\widetilde{V}%
_{q^{\prime }}^{m^{\prime \prime }}\overline W_{\Lambda }^{(2)}\overline\Pi_\Lambda ^{-1}]_{m^{\prime
\prime }m^{\prime \prime }}{\overline G_\Lambda}\right.  \nonumber \\
&&\times \left. \lbrack \overline \Pi_\Lambda ^{-1}\overline W_{\Lambda ,q^{\prime }-q}^{(2)}\overline \Pi_\Lambda
^{-1}-\overline \Pi_\Lambda ^{-1}]_{m^{\prime \prime \prime }m^{\prime \prime \prime
}}{\overline G_\Lambda}\right\} _{k^{\prime \prime }k^{\prime \prime \prime }}-
\widetilde{%
V}_{q}^{m}\delta _{mm^{\prime }}
\nonumber
\\
&&+\sum_{m^{\prime \prime }=c,s}a_{m^{\prime \prime }}^{(m)}[\overline\Pi_\Lambda ^{-1}\overline W_{\Lambda
}^{(2)}\widetilde{V}_{k^{\prime }-k}^{m^{\prime \prime }}\overline W_{\Lambda
}^{(2)}\overline \Pi_\Lambda ^{-1}]_{k,k+q,k^{\prime }-k,m^{\prime \prime }m^{\prime \prime
}}\delta _{mm^{\prime }}\notag \\&&+\sum\limits_{k'',m''}\overline{\Phi }_{\Lambda,kk^{\prime }q;k''}^{(3),mm'm''}%
\frac{{d}\overline{G}^{m''}_{\Lambda ,k''}}{{d}\Lambda },  \label{G2}
\end{eqnarray}%
where last term in each equation appears because of the dependence of stationary Green function on $\Lambda$ (cf. Ref. \cite{Dupuis1}); the coeffitients $a_{m^{\prime }}^{(m)}$, are given after the Eqs. (\ref{flow_eq}) of
the main text, and $c_{m^{\prime \prime }m^{\prime \prime \prime
}}^{(m,m^{\prime })}$ are some coefficients. In the main text of the paper we neglect ``Aslamazov-Larkin''
contribution (first term in the r.h.s. of Eq. (\ref{G2})), which is a rather common approximation (cf. Ref. \cite{Bickers1}). 
Second terms in the right hand sides of the Eqs. (\ref{G1}) and  (\ref{G2}) can be removed by representing 
\begin{eqnarray}
\overline\Phi _{\Lambda,k}^{(1),m} &=&\widetilde\Phi _{\Lambda,k}^{(1),m}-\Lambda \widetilde{%
V}_{q=0}^{m}\sum_{k^{\prime }} {\overline G}_{\Lambda,k^{\prime
}}^{m}\notag \\
\overline\Phi _{\Lambda,kk^{\prime }q}^{(2),mm^{\prime }} &=&\widetilde\Phi _{\Lambda,kk^{\prime }q}^{(2),mm^{\prime }}-\Lambda \widetilde{%
V}_{q}^{m}\delta _{mm^{\prime }}
\end{eqnarray}
which implies also a change $\overline\Phi _{\Lambda,kk^{\prime }q}^{(2),mm^{\prime }} \rightarrow\widetilde\Phi _{\Lambda,kk^{\prime }q}^{(2),mm^{\prime }}$ in the last term in the right-hand side of Eq. (\ref{G1}).
Representing $\overline{\Phi }^{(3)}$ as all
possible combinations of ${\mathcal F}^{m}
G_{\Lambda }{\mathcal F}^{m}$ (which generalizes the representation of this vertex as combinations of $\Phi^{(2)}G\Phi^{(2)}$ in Refs. \cite{Dupuis1,Meden}), we obtain Eqs. (\ref{flow_eq}) of the main
text. Note that in the considered approximation the three-particle vertex in Eq. (\ref{G2}) corresponds to the two-particle contribution to $\Phi^{(2)}$, in contrast to the terms, neglected in the Eqs. (\ref{A5}) and (\ref{A6}).

\section{Relation to the one-particle-irreducible approach for
vanishing non-local interaction}

In this Appendix we consider the relation of the 2PI-fRG equations (\ref{flow_eq}) for
vanishing non-local interaction to the equations of the DMF$^{2}$RG
approach. More generally, this concerns the relation between
2PI and 1PI fRG approaches. The equations of the latter approach can be
written in the form (see, e.g., Ref. \cite{fRGReview}; for the diagrammatic representation see Fig. 4
of that paper)%
\begin{subequations}
\begin{eqnarray}
\frac{d\Sigma _{\Lambda,1}}{d\Lambda } &=&{ F} _{\Lambda,12,12}S_{\Lambda,2} \label{fRGEq1}\\
\frac{d{ F} _{\Lambda,12;1^{\prime }2^{\prime }}}{d\Lambda } &=&-\frac{1}{2}{ F}
_{\Lambda,12,1^{\prime \prime }2^{\prime \prime }}\frac{dP_{\Lambda,{1''2''}}}{d\Lambda} { F} _{\Lambda,1^{\prime \prime }2^{\prime \prime
},1^{\prime }2^{\prime }}+{ F} _{\Lambda,12^{\prime \prime },1^{\prime }1^{\prime
\prime }}\frac{dP_{\Lambda,{1''2''}}}{d\Lambda }{ F} _{\Lambda,1^{\prime \prime }2,2^{\prime \prime }2^{\prime }}\notag\\
&&-{ F}
_{\Lambda,12^{\prime \prime },2^{\prime }1^{\prime \prime }}\frac{dP_{\Lambda,{\bf 1''2''}}}{d\Lambda }%
{ F} _{\Lambda,1^{\prime \prime
}2,2^{\prime \prime }1^{\prime }}+{ F} _{\Lambda,121^{\prime \prime },1^{\prime
}2^{\prime }1^{\prime \prime }}^{(6)}S_{\Lambda,1^{\prime \prime }},\label{fRGEq2}
\end{eqnarray}%
\end{subequations}
where ${F}_{\Lambda,12;1'2'}$ and ${F}^{(6)}_{\Lambda,123;1'2'3'}$ are the two- and three-particle 1PI interaction vertices, respectively, indexes $1=(\sigma _{1},k_{1})$ etc. denote spin-, and
frequency-momenum variables, the first pair of indexes $(1,2)$ in the vertex ${F}_{\Lambda,12;1'2'}$ corresponds to incoming, and second pair $(1',2')$ to the outgoing particles, we assume spin-, momentum- and frequency
conservation in the vertices, and $P_{\Lambda,{1''2''}}=G_{\Lambda,{1''}}G_{\Lambda,{2''}}$. The Green functions are assumed spin
independent, 
the``single-scale" propagator $S_{\Lambda ,k}=-G_{\Lambda,k}^2 (d{G}_{0\Lambda ,k}^{-1}/d\Lambda)$,
and we have performed the replacement \cite{Katanintrunc} $S\rightarrow dG/d\Lambda$ in the equation (\ref{fRGEq2}). 

The equation (\ref{fRGEq1}) can be put easily in the 2PI form by accounting for spin independence of $S_\Lambda$ and introducing 
$F ^{c(s)}_{\Lambda,{\bf 12;1'2'}}=-F_{\Lambda,\bf 12;1'2'}
^{\uparrow \uparrow }\mp F_{\Lambda,\bf 12;1'2'} ^{\uparrow \downarrow }$, where $F _{\Lambda,\mathbf{12};%
\mathbf{1}^{\prime }\mathbf{2}^{\prime }}^{\uparrow \uparrow, \uparrow \downarrow }$  denote vertices $F
_{\Lambda,12;1^{\prime }2^{\prime }}$ with $\sigma _{1}=\sigma _{1}^{\prime
}=\uparrow $ and $\sigma _{2}=\sigma _{2^{\prime }}=\uparrow ,\downarrow$, respecively ($\mathbf{1}=k_{1}$ etc.). Representing 
\begin{eqnarray}
F^{c(s)}_{\Lambda,\bf 12;12} &=&2[1+\Phi ^{(2),c(s)}_{\Lambda,\bf 11',0} \Pi_{\Lambda,{\bf 1'},0} ]^{-1}_{\bf 1,1'}\Phi ^{(2),c(s)}_{\Lambda,\bf 1'2,0},\label{BS}\\ S_{\Lambda,\bf 1}&=&dG_{\Lambda,\bf 1}/d\Lambda-(\Pi_{\Lambda,{\bf 1},0}/2) (d\Sigma_{\Lambda,\bf 1}/d\Lambda), 
\end{eqnarray}
where 
$\Pi_{\Lambda,k,0}=2G_{\Lambda,k}^2$, we obtain the equation
\begin{equation}
\dot{\Sigma}_{\Lambda ,\bf 1}=-2\Phi _{\Lambda
,{\bf 1,2},0}^{(2),c} \frac{{d}\overline{G}_{\Lambda ,\bf 2}}{%
{d}\Lambda },  \label{dS}
\end{equation}%
which is identical to the equation (\ref{Sflow_eq}) for $V=\widetilde{V}=$ $R=0$. 

To outline the derivation of equations for 2PI vertices, we
consider for concreteness charge and spin channels. By combining equations for $F _{\Lambda,\mathbf{12};%
\mathbf{1}^{\prime }\mathbf{2}^{\prime }}^{\uparrow \uparrow }$ and $F
_{\Lambda,\mathbf{12};\mathbf{1}^{\prime }\mathbf{2}^{\prime }}^{\uparrow \downarrow
}$ and introducing in addition to charge- and spin vertices singlet and triplet components, defined by $F^{\rm ssc(tsc)}_{\Lambda,\mathbf{12};\mathbf{1}^{\prime }\mathbf{2}^{\prime }}=(\Gamma
_{\Lambda,\mathbf{12};\mathbf{1}^{\prime }\mathbf{2}^{\prime }}^{ \uparrow \downarrow
}\mp \Gamma
_{\Lambda,\mathbf{12};\mathbf{1}^{\prime }\mathbf{2}^{\prime }}^{\downarrow \uparrow
})/2$, where $F
_{\Lambda,\mathbf{12};\mathbf{1}^{\prime }\mathbf{2}^{\prime }}^{\downarrow \uparrow
}=F
_{\Lambda,12;1^{\prime }2^{\prime }}$ with $\sigma _{1}=\sigma _{2}^{\prime
}=\uparrow $ and $\sigma _{2}=\sigma _{1^{\prime }}=\downarrow$, we find%
\begin{eqnarray}
\frac{dF _{\Lambda,\mathbf{12};\mathbf{1}^{\prime }\mathbf{2}^{\prime }}^{c(s)}}{%
d\Lambda } &=&\sum_{m={\rm ssc,tsc}} b^{(c(s))}_{m} F _{\Lambda,\mathbf{12},\mathbf{1}^{\prime
\prime }\mathbf{2}^{\prime \prime }}^{m }\frac{{d}%
P_{\Lambda,{\bf 1''2''}}}{d\Lambda }F _{%
\Lambda,\mathbf{1}^{\prime \prime }\mathbf{2}^{\prime \prime },\mathbf{1}^{\prime }%
\mathbf{2}^{\prime }}^{m }  -F _{\Lambda,\mathbf{12}^{\prime \prime },\mathbf{1}^{\prime }\mathbf{1}%
^{\prime \prime }}^{c(s)}\frac{dP_{\Lambda,{\bf 1''2''}}}{d\Lambda }F _{\Lambda,\mathbf{1}^{\prime \prime }%
\mathbf{2},\mathbf{2}^{\prime \prime }\mathbf{2}^{\prime
}}^{c(s)}\notag\\
&&+\sum\limits_{m=c,s}a_{m}^{(c(s))}F _{\Lambda,\mathbf{12}^{\prime \prime
},\mathbf{2}^{\prime }\mathbf{1}^{\prime \prime }}^{m}\frac{dP_{\Lambda,{\bf 1''2''}}}{d\Lambda }F _{\Lambda,\mathbf{%
1}^{\prime \prime }\mathbf{2},\mathbf{2}^{\prime \prime }\mathbf{1}^{\prime
}}^{m} +F ^{(6)}\ast S\label{dGamma}
\end{eqnarray}%
where $a_{m}^{(c,s)}$ are defined after Eqs. (\ref{flow_eq}) of the paper, $*$ stands for summation over respective spin-, momenta-, and frequency indexes, and we have accounted that due to $%
SU(2)$ symmetry $F _{\mathbf{12};\mathbf{1}^{\prime }\mathbf{2}^{\prime
}}^{\uparrow \downarrow }=-F _{\mathbf{12};\mathbf{2}^{\prime }\mathbf{1}%
^{\prime }}^{s}$, $F
_{\Lambda,\mathbf{12};\mathbf{1}^{\prime }\mathbf{2}^{\prime }}^{ \uparrow \uparrow
}=2F^{\rm tr}_{\Lambda,\mathbf{12};\mathbf{1}^{\prime }\mathbf{2}^{\prime }}$. On the next step we use again equation (\ref{BS}).
By differentiating this equation we obtain%
\begin{eqnarray}
\frac{dF _{\Lambda,\mathbf{12};\mathbf{1}^{\prime }\mathbf{2}^{\prime }}^{c(s)}}{%
d\Lambda } &=
&-\frac{1}{2}\left[F_\Lambda ^{c(s)}\overset{.}{\Pi }_\Lambda F_\Lambda ^{c(s)}\right]_{\bf 11';22'}\notag\\&+&2\left\{[1+\Phi_\Lambda ^{c(s)}\Pi_\Lambda ]^{-1}%
\overset{.}{\Phi }_\Lambda^{c(s)}[1+\Pi_\Lambda \Phi ^{c(s)}_\Lambda]^{-1}\right\}_{\bf 11';22'}
\label{dBSEq}
\end{eqnarray}%
where matrix multiplication and inversion with respect to specified groups of indexes is assumed; $\Pi_\Lambda$ is considered as diagonal matrix. Combining Eqs. (\ref{dGamma}) and (\ref{dBSEq}) we obtain%
\begin{eqnarray}
\frac{d\Phi _{\Lambda,\mathbf{12};\mathbf{1}^{\prime }\mathbf{2}^{\prime }}^{c(s)}}{%
d\Lambda } &=&\frac{1}{2}\left[1+\Phi_\Lambda ^{c,s}\Pi_\Lambda \right]_{{\bf 1}{\tilde {\bf 1}};{\bf 2}\tilde{\bf 2}}\left\{ \sum_{m={\rm ssc,tsc}} b^{(c(s))}_{m}F _{\Lambda,{\tilde{\bf 1}\tilde{\bf 2}},\mathbf{1}^{\prime
\prime }\mathbf{2}^{\prime \prime }}^{m }\frac{{d}%
P_{\Lambda,{\bf 1''2''}}}{d\Lambda }F _{%
\Lambda,\mathbf{1}^{\prime \prime }\mathbf{2}^{\prime \prime },\tilde{\mathbf{1}}^{\prime }%
\tilde{\mathbf{2}}^{\prime }}^{m } \right.  \\
&&\left. +\sum\limits_{m=c,s}a_{m}^{(c,s)}F _{\Lambda,{\tilde{\bf 1}{\bf 2}}^{\prime
\prime },\tilde{\mathbf{2}}^{\prime }\mathbf{1}^{\prime \prime }}^{m}\frac{dP_{\Lambda,{\bf 1''2''}}}{%
d\Lambda }F _{\Lambda,\mathbf{1}^{\prime \prime }\tilde{\mathbf{2}},\mathbf{2}^{\prime \prime }%
\tilde{\mathbf{1}}^{\prime }}^{m}+F ^{(6)}\ast S\right\} [1+\Pi _\Lambda \Phi_\Lambda ^{c,s}]_{\tilde {\bf 1}'{\bf 1'};\tilde{\bf 2}'{\bf 2}'}\notag
\end{eqnarray}%
The factors $[1+\Phi_\Lambda ^{c,s}\Pi_\Lambda ]$ remove the
two-particle reducible contributions. However, such contributions can be
generated by the three-particle vertex term $F ^{(6)}\ast S$ only; when
this term is neglected the mentioned factors remove the diagrams which are not added. This
situation is similar to the one appearing in the dual fermion approach \cite{MySixpt} where
the self-energy acquires spurious denominator, which does not have any
diagrammatic representation, due to neglect of the three-particle vertices.
Therefore, at the two-particle level it is consistent to omit these
factors when neglecting $F ^{(6)};$ the final equation for the
two-particle vertex therefore reads 
\begin{eqnarray}
\frac{d\Phi _{\Lambda,\mathbf{12};\mathbf{1}^{\prime }\mathbf{2}^{\prime }}^{c,s}}{%
d\Lambda } &=&\frac{1}{2}\left\{ \sum_{m={\rm ssc,tsc}} b^{(c(s))}_{m}F _{\Lambda,\mathbf{12},\mathbf{1}^{\prime
\prime }\mathbf{2}^{\prime \prime }}^{m }\frac{{d}%
P_{\Lambda,{\bf 1''2''}}}{d\Lambda }F _{%
\Lambda,\mathbf{1}^{\prime \prime }\mathbf{2}^{\prime \prime },\mathbf{1}^{\prime }%
\mathbf{2}^{\prime }}^{m } \right.  \notag\\
&&\left. +\sum\limits_{m=c,s}a_{m}^{(c,s)}F _{\Lambda,\mathbf{12}^{\prime
\prime },\mathbf{2}^{\prime }\mathbf{1}^{\prime \prime }}^{m}\frac{dP_{\Lambda,{\bf 1''2''}}}{%
d\Lambda }F _{\Lambda,\mathbf{1}^{\prime \prime }\mathbf{2},\mathbf{2}^{\prime \prime }%
\mathbf{1}^{\prime }}^{m}\right\} 
\end{eqnarray}%
and it is consistent with the equation (\ref{ifflow_eq}) of the paper, if we take into account that $F _{\Lambda,\mathbf{12},\mathbf{1}^{\prime
}\mathbf{2}^{ \prime }}^{\rm ssc(tsc)}={\mathcal F} _{\Lambda,\mathbf{1},\mathbf{1}^{\prime
 },{\bf 1}+\mathbf{2}}^{\rm ssc(tsc)}$, $F_{\Lambda,\mathbf{12},\mathbf{1}^{\prime
}\mathbf{2}^{ \prime }}^{\rm s(c)}=2{\mathcal F} _{\Lambda,\mathbf{1},\mathbf{1}^{\prime
 },{\bf 2'}-\mathbf{1}}^{\rm s(c)}$, where the vertices in the right-hand sides refer to those used in the main text of the paper.

\section{The equivalence of the susceptibility (\protect\ref%
{CDBL}) to the result of the dual boson approach}

The physical charge or spin susceptibility in the dual boson approach is
given by \cite{DB3}%
\begin{equation}
X_{{q}}=\frac{1}{1/\Pi _{{q}}^{(1)}+V_{\mathbf{q}%
}-v(\omega )},
\end{equation}%
where

\[
\Pi _{{q}}^{(1)}=2\int d\nu ^{\prime \prime }\int d\nu
^{\prime }\chi _{\nu ^{\prime  },q }^{0}\left[
\delta_{\nu'\nu''}-\Phi _{\mathrm{loc,}\nu ^{\prime }\nu ^{ \prime \prime
}\omega }^{(2)}\chi _{\nu ^{ \prime \prime},q  }^{0}%
\right] _{\nu ^{ \prime }\nu ^{ \prime \prime }}^{-1}, 
\]%
$\chi _{\nu,q }^{0}=-2\sum\nolimits_{\mathbf{k}}\overline{G}%
_{1,\mathbf{k}\nu }\overline{G}_{1,\mathbf{k+q,}\nu +\omega }$ is the $%
\Lambda =1$ limit of the Eq. (\ref{chi0}), and we consider here only one
specific channel (charge or spin). Now we introduce the quantities 
\begin{eqnarray}
\Phi _{\nu \nu ^{\prime }{q} } &=&[(\chi _{\nu,q}^{0})^{-1}\delta_{\nu\nu'}-\Phi _{\mathrm{loc,}\nu \nu ' \omega }^{(2)}+\tilde U_{{q}%
}]^{-1}=\chi _{\nu q }^{0}[\delta_{\nu \nu'}-(\Phi _{\mathrm{loc,}\nu
\nu ' \omega }^{(2)}
-\tilde U_{%
q })\chi _{\nu ^{\prime },q }^{0}]^{-1},  \nonumber \\
\phi _{{q} } &=&\int d\nu d\nu ^{\prime }\Phi _{\nu
\nu ^{\prime }q }
\end{eqnarray}%
with some $\tilde U_{q }.$ Then we obtain%
\begin{eqnarray}
\Pi _{q }^{(1)} &=&\int d\nu d\nu ^{\prime }d\nu ^{\prime
\prime }d\nu ^{\prime \prime \prime }\chi _{\nu, q }^{0}\left[
\delta_{\nu \nu'}-\Phi _{\mathrm{loc,}\nu \nu \prime \omega }^{(2)}\chi _{\nu
^{\prime }q }^{0}+\tilde U_{q }\chi _{\nu ^{\prime
},q }^{0}\right] _{\nu \nu ^{\prime }}^{-1}  \nonumber \\
&&\times \left[ \delta_{\nu' \nu''}-\Phi _{\mathrm{loc,}\nu ^{\prime }\nu ^{\prime \prime
}\omega }^{(2)}\chi _{\nu ^{\prime \prime },q }^{0}+\tilde U_{q }\chi _{\nu ^{\prime \prime },q }^{0}\right] \left[
\delta_{\nu'' \nu'''}-\Phi _{\mathrm{loc,}\nu ^{\prime \prime }\nu ^{\prime \prime \prime
}\omega }^{(2)}\chi _{\nu ^{\prime \prime \prime },q }^{0}%
\right] _{\nu ^{\prime \prime }\nu ^{\prime \prime \prime }}^{-1}  \nonumber
\\
&=&\int d\nu d\nu ^{\prime }\Phi _{\nu \nu ^{\prime }q
}\left\{ 1+\tilde U_{q }\int d\nu ^{\prime \prime }d\nu ^{\prime
\prime \prime }\chi _{\nu ^{\prime \prime },q }^{0}\left[
\delta_{\nu'' \nu'''}-\Phi _{\mathrm{loc,}\nu ^{\prime \prime }\nu ^{\prime \prime \prime
}\omega }^{(2)}\chi _{\nu ^{\prime \prime \prime },q }^{0}%
\right] _{\nu ^{\prime \prime }\nu ^{\prime \prime \prime }}^{-1}\right\} 
\nonumber \\
&=&\phi _{q }(1+\tilde U_{q }\Pi _{q }^{(1)}).
\end{eqnarray}%
Therefore,%
\begin{equation}
\Pi _{q }^{(1)}=\frac{1}{\phi _{q }^{-1}-\tilde U_{%
q }},
\end{equation}%
and%
\begin{equation}
X_{q }=\frac{1}{\phi _{q }^{-1}-\tilde U_{q }+V_{\mathbf{q}}-v(\omega )}.
\end{equation}%
The choice $\tilde U_{q }=V_{\mathbf{q}}-v(\omega )$ yields $X_{%
q }=\phi _{q }=\int d\nu d\nu ^{\prime }[(\chi
_{\nu,q }^{0})^{-1}\delta _{\nu \nu ^{\prime }}-\Phi _{%
\mathrm{loc,}\nu \nu ^{\prime }\omega }^{(2)}+V_{\mathbf{q}}-v(\omega
)]_{\nu \nu ^{\prime }}^{-1},$ which is equivalent to the Eq. (\ref{CDBL})
of the main text at $\Lambda =1$. On the other hand, choosing $U_{q }=\mp U/2-v(\omega)$ leads us to the $\Lambda =1$ limit of the Eqs.
 (\ref{chiL}) and  (\ref{phiL}) of the main text for the local 2PI interaction $\tilde \Phi _{\Lambda,k k' q
}^{(2),c(s)}= \Phi _{{\rm loc},k k' q
}^{(2),c(s)}$. Analogously one can prove the equivalence of Eqs. (\ref{CDBL1}) and (\ref{chiL}) for arbitrary $\Lambda$ and non-local 2PI interaction, by replacing integrals over frequencies by the corresponding momenta-frequency sums and appropriately choosing $\tilde U_q$.


\begin{thebibliography}{99}
\bibitem{Note} We note that in Ref. \cite{DB2} the corresponding condition for instability in RPA was incorrectly written as $V^c_{\rm RPA}=U/4+{\rm const}$. This yielded incorrect slope of phase boundary in RPA (see discussion of the phase diagram). At the same time, numerical results obtained in the dual boson approximation in the same reference agree with the present analysis.
\end{thebibliography}
\end{document}